%% file: main.tex
\documentclass{sig-alternate}
\usepackage{mathptmx} %

\usepackage{fancyhdr}
\usepackage[normalem]{ulem}
\usepackage[hyphens]{url}
\usepackage[sort,nocompress]{cite}
\usepackage[final]{microtype}
\usepackage[keeplastbox]{flushend}
\usepackage[bookmarks=true,breaklinks=true,letterpaper=true,colorlinks,citecolor=blue,linkcolor=blue,urlcolor=blue]{hyperref}

\include{commands}
\title{\titleName}

\author{%
	Anbang Wu \\
	Department of Computer Science\\
	University of California, Santa Barbara \\
	anbang@ucsb.edu
	\and
	Hezi Zhang \\
	Department of Computer Science \\
	University of California, Santa Barbara \\
	hezi@ucsb.edu
	\and
	Gushu Li \\
	Department of Electrical \& Computer Engineering\\
	University of California, Santa Barbara \\
	gushuli@ece.ucsb.edu
	\and
	Alireza Shabani \\
	Cisco Research \\
	Los Angeles, California \\
	ashabani@cisco.com
	\and
	Yuan Xie \\
	Department of Electrical \& Computer Engineering\\
	University of California, Santa Barbara \\
	yuanxie@ucsb.edu
	\and
	Yufei Ding \\
	Department of Computer Science\\
	University of California, Santa Barbara \\
	yufeiding@cs.ucsb.edu
}

\begin{document}
\maketitle
\pagestyle{plain}

\input{01_abstract}

\input{02_introduction}

\input{04_background}

\input{05_algorithm}

\input{07_evaluation}

\input{07_discussion}

\input{03_related_work}

\input{08_conclusion}

\bibliographystyle{unsrt}
\bibliography{references}

\end{document}

%% file: commands.tex
\usepackage{amsmath,mathtools,amssymb}

\usepackage{titlesec}
\usepackage{multirow}
\usepackage{tikz}
\usepackage{adjustbox}
\usetikzlibrary{quantikz}
\DeclareUnicodeCharacter{2009}{\,} 
\titlespacing*{\section}
{0pt}{\parsep}{\parsep}
\titlespacing*{\subsection}
{0pt}{\parsep}{\parsep}
\usepackage[ruled,vlined,linesnumbered]{algorithm2e}

\newcommand\circlenum[1]{\raisebox{.5pt}{\textcircled{\raisebox{-.9pt} {#1}}}}

\def\frameworkName {AutoComm}
\def\frameworkNameSpace {AutoComm }
\def\titleName {\frameworkName: A Framework for Enabling Efficient Communication in Distributed Quantum Programs}

%% file: 01_abstract.tex
\begin{abstract}
Distributed quantum computing (DQC) is a promising approach to extending the computational power of near-term quantum devices. 
However, the non-local quantum communication between quantum devices is much more expensive and error-prone than
the local quantum communication within each quantum device.
Previous work on the DQC communication optimization focus on optimizing the communication protocol for each individual non-local gate and then adopt quantum compilation designs which are designed for local multi-qubit gates (such as controlled-x or CX gates) in a single quantum computer.
The communication patterns in distributed quantum programs are not yet well studied, leading to a far-from-optimal communication cost.
In this paper, we identify \textit{burst communication}, a specific qubit-node communication pattern that widely exists in many distributed programs and can be leveraged to guide communication overhead optimization.
We then propose \frameworkName, an automatic compiler framework to first extract the burst communication patterns from the input programs, and then optimize the communication steps of burst communication discovered.
Experimental results show that our proposed \frameworkNameSpace can reduce the communication resource consumption and the program latency by 75.6\% and 71.4\% on average, respectively.

\end{abstract}

%% file: 02_introduction.tex
\section{Introduction}

Quantum computing is promising with its great potential of providing significant speedup to many problems, such as large-number factorization with an exponential speedup \cite{Shor} and unordered database search with a quadratic speedup \cite{Grover}.
A large number of qubits is required in order to solve practical problems with quantum advantage and the qubit count requirement is even higher after taking quantum error correction \cite{nielsen2002quantum} into consideration. 
However, it has turned out that extending the number of qubits on a single quantum processor is exceedingly difficult due to various hardware-level challenges such as crosstalk errors \cite{crosstalk1,crosstalk2}, qubit addressability~\cite{bruzewicz2019trapped}, fabrication difficulty~\cite{brink2018device}, etc.
The challenges usually increase with the size of quantum hardware and may limit the number of qubits accommodated by a single quantum processor.  %

Rather than relying on the advancement of a single quantum processor, an alternative way of increasing scalability is by distributed quantum computing (DQC), which integrates the computing resources of multiple modular quantum processors. %
For example,  recent experiments have demonstrated an entanglement-based quantum network of three quantum processors \cite{QNexp_3-node_network}. Companies such as IBM also envision in their roadmap \cite{IBM2020roadmap} a
future of creating a large-scale quantum computer
with quantum interconnects that link superconducting quantum processors.
Similarly, the ion trap-based quantum computer also 
requires an optical network of multiple traps each with tens of qubits in-order to scale up, making DQC a path to realizing large-scale quantum computers  \cite{monroe2012}.

In DQC, remote communication involving qubits in different computing nodes is essential yet far more expensive than the local communication on qubits within the same node (e.g., 5-100x time consumption and up to 40x accuracy degradation~\cite{time-slice, Young2022AnAF}).
There are two major schemes for remote quantum communication: one built upon the cat-entangler and cat-disentangler protocol~\cite{cat_entangle}, and the other based on the quantum teleportation~\cite{nielsen2002quantum}. 
In this paper, we denote the former scheme as Cat-Comm and the latter one as TP-Comm. 
Both schemes consume EPR pairs~\cite{optimal_nonlocal}, which are pre-distributed entangled qubit pairs, as a resource to establish quantum communication. %
Cat-Comm can implement the remote CX gate~\cite{nielsen2002quantum} with only one EPR pair, but for  general two-qubit gates like the SWAP gate~\cite{nielsen2002quantum}, Cat-Comm  requires up to three EPR pairs~\cite{Ferrari2021CompilerDF}. In contrast, TP-Comm conducts any remote two-qubit gate with two EPR pairs~\cite{optimal_nonlocal}, making it more efficient for the SWAP gate. For a distributed program,  more complex remote operations or more information getting transferred per EPR pair would lead to less communication cost.

The overall compiling flow for DQC is similar to that of single-node quantum programs, except with more emphasis on remote communication overhead. 
Ferrari et al.~\cite{Ferrari2021CompilerDF} propose a compiler design similar to single-node compilers~~\cite{Li2019TacklingTQ, Qiskit, Amy2019staqAFQ, Khammassi2022OpenQLA, Sivarajah2020tketAR} using Cat-Comm for each remote CX gate and TP-Comm for each remote SWAP gate. Unsatisfied with the low information of the remote CX gate, Baker et al.~\cite{time-slice} eliminate all remote CX gates by using the remote SWAP gate, which only requires two EPR pairs for implementation but contains the information of three CX gates. Unfortunately, bounded by the information of a single two-qubit gate, these compilers cannot achieve higher throughput of information per EPR pair.

Eisert et al.~\cite{optimal_nonlocal} suggest higher throughput could be achieved by considering multi-qubit gates. Diadamo et al.~\cite{Diadamo2021DistributedQC} propose a specialized compiler for distributed VQE that uses Cat-Comm to implement controlled-unitary-unitary and controlled-controlled-unitary gates. However, their work can only optimize the gate written in the controlled-unitary form and thus cannot work with decomposed circuits. Moreover, their work
cannot optimize programs lacking controlled-unitary blocks.

Besides increasing the `height' (number of qubits) of remote operations, we observe that the throughput of information per EPR pair can also be significantly boosted up by expanding the `width'  (number of gates) of each remote communication. Specifically, we discover that a large amount of remote two-qubit gates in distributed quantum programs can be implemented collectively through one or two communication invocations. On top of the observation, we propose to optimize the communication overhead based on the \textit{burst communication}, which denotes a group of continuous remote two-qubit gates between one qubit and one node. Burst communication is powerful as it is more information-intensive than single two-qubit gate and contains but not limited to controlled-unitary blocks. Burst communication is also flexible for optimization as it does not require specialized circuit representation and is available in decomposed circuits.

\begin{figure}
    \centering
    \includegraphics[width=0.47\textwidth]{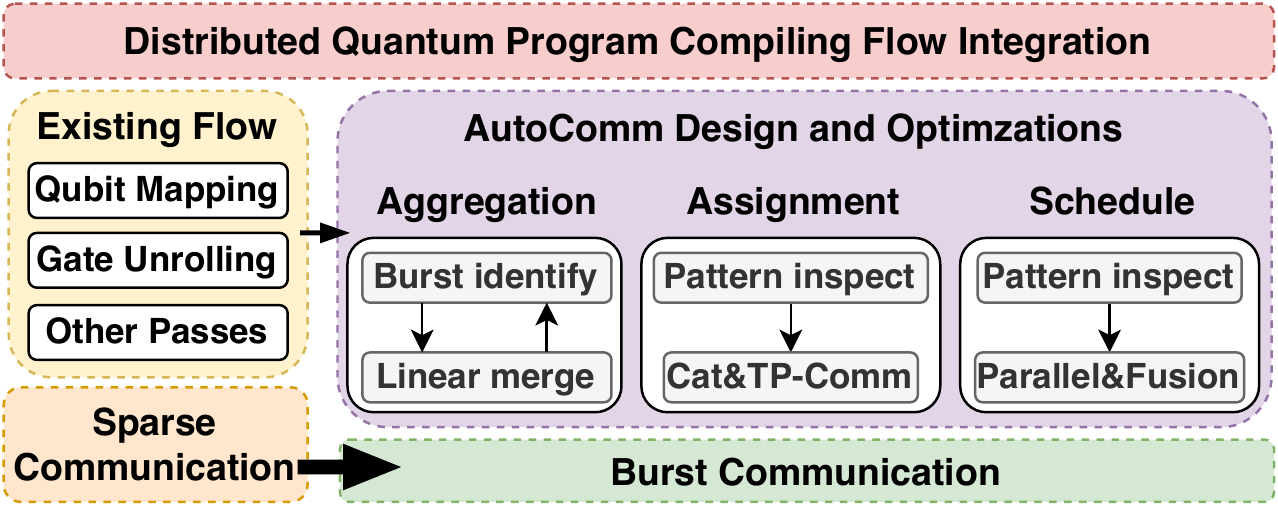}
    \caption{\frameworkNameSpace Overview.}
    \label{fig:designoverview}
\end{figure}

To this end, we develop the first burst-communication-centric optimization framework, \textit{\frameworkName} as shown in Figure~\ref{fig:designoverview}.
In contrast to existing compiling flows~\cite{time-slice, Ferrari2021CompilerDF, Li2019TacklingTQ, Qiskit, Amy2019staqAFQ, Khammassi2022OpenQLA, Sivarajah2020tketAR}, where each remote CX gate is implemented independently (i.e., sparse communication),  \textit{\frameworkName} greatly mitigates the communication bottleneck with burst communication and can be easily integrated into these existing compiling flows.
Our framework consists of three key stages. Firstly, we perform a communication aggregation pass to group remote gates and extract burst communication blocks. 
Due to the broad availability of burst communication in distributed quantum programs, this pass could generate a large amount of burst communication blocks for following optimizations. 
Secondly, we propose a hybrid communication scheme which examines the patterns of each burst communication block and assigns the optimal communication scheme for each block. The insight for this step is that, TP-Comm and Cat-Comm is more resource-efficient for different type of burst communications and considering only one communication scheme would incur extra resource consumption.
Finally, we propose an adaptive communication schedule for burst communication blocks of different patterns to squeeze out the parallelism between them and thus reduce overall program latency.
There are two critical observations for this optimization: it is possible to execute burst communication with shared qubits or nodes in parallel, and we can fuse some burst communication blocks to cut down the communication footprint.

Our contributions are summarized as follows:
\begin{itemize}
    \item We identify the burst communication feature in distributed quantum computing and promote its importance in optimizing distributed quantum programs. We further propose the first communication optimization framework based on the burst communication. %
    \item We propose a communication aggregation pass to expose burst communications of distributed quantum programs and then design a hybrid communication scheme, using both Cat-Comm and TP-Comm to accommodate different communication patterns.
    \item We propose an efficient communication scheduling method to optimize the latency adaptively squeezing out the parallelism of various patterns. %
    \item  Compared to the state-of-the-art baseline method~\cite{Ferrari2021CompilerDF}, \frameworkNameSpace significantly reduces the communication resource consumption and the program latency by 75.6\% and 71.4\% on average, respectively. 

    \end{itemize}

%% file: 04_background.tex
\section{Background} \label{sect: bg}

In this section, we introduce necessary background to understand the  distributed quantum computing and its communication.
We do not cover the basic quantum computing concepts (e.g., qubit, gate, measurement) and recommend ~\cite{nielsen2002quantum} for more details. %

\subsection{EPR Pair and Entanglement}

\paragraph{EPR entanglement} To establish quantum communication in a distributed quantum computer, we first need to generate a pair of qubits whose state is $\frac{1}{\sqrt{2}}(\ket{00} + \ket{11})$, EPR entangled state. 
The two qubits such state is called EPR entanglement pair (Abbrev., EPR pair)~\cite{nielsen2002quantum}. 
The two qubits of an EPR pair can be distributed on different quantum devices, formulating a remote EPR pair~\cite{optimal_nonlocal}. 
The preparation of the remote EPR pair includes two stages: generation and purification. The generation stage generates and distributes EPR pairs but is very noisy, making the purification stage indispensable~\cite{isailovic2006interconnection}.

\subsection{Distributed Quantum Computing}

The development of quantum communication~\cite{QNexp9, QNexp10, QNexp11, QNexp12, QNexp13, QNexp14, QNexp15, QNexp16, QNexp17, QNexp18, QNexp19, QNexp_3-node_network} enables distributed computing over a series of quantum devices. As in classical distributed computing, remote communication between computing nodes is also the bottleneck of distributed quantum computing (DQC) and should be carefully optimized.

\begin{figure}[h!] \small
    \centering
    \includegraphics[height=0.16\textwidth]{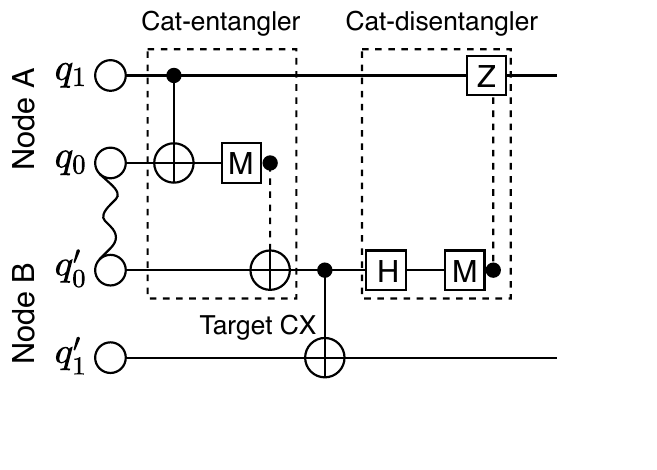}\hfill{} \includegraphics[height=0.16\textwidth]{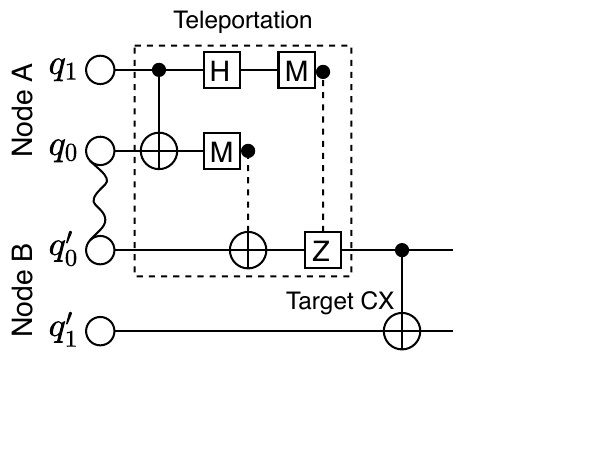} \\
    \hphantom{}\hspace{30pt}(a)\hspace{100pt}(b)
    \caption{The implementation of one remote CX. (a) The Cat-Comm version. (b) The TP-Comm version. Each wavy line denotes an EPR pair between qubits, and each dashed line denotes one bit of classical communication. M denotes measurement.}
    \label{fig:qucomm}
\end{figure}

Different from the classical distributed computing system, quantum data cannot be easily shared across quantum nodes due to the no-cloning theorem~\cite{nielsen2002quantum}. 
The workaround is to exploit different  communication schemes (e.g., \textit{Cat-Comm}~\cite{optimal_nonlocal} and \textit{TP-Comm}~\cite{nielsen2002quantum}) based on remote EPR entanglement, one of the key information resources in quantum processing.
Figure~\ref{fig:qucomm} illustrates how to use these two schemes to implement one \textit{remote CX gate}, with the control qubit $q_1$ residing in quantum nodes A and the target qubit $q_1'$ in node B. Qubits in Figure~\ref{fig:qucomm} fall into two categories. The first category of qubits is used to store quantum information and is called \textit{data qubits}, e.g., $q_1$ and $q_1'$. The second category of qubits, called \textit{communication qubits}, is used to hold the remote EPR entanglement required for quantum communication, e.g. $q_0$ and $q_0'$ in Figure~\ref{fig:qucomm}.

As shown in Figure~\ref{fig:qucomm}(a), the first communication scheme Cat-Comm utilizes cat-entangler to transfer the state of the control qubit $q_1$ to node B, execute the target CX gate, and then use cat-disentangler to transfer the state back to node A. While TP-Comm, the second communication scheme in Figure~\ref{fig:qucomm}(b), employs quantum teleportation~\cite{nielsen2002quantum} to transfer the state of $q_1$, and then execute the target CX gate. Though Cat-Comm and TP-Comm both require one EPR pair and two bits of classical communication,
Cat-Comm is more widely-used than TP-Comm in DQC compilers~\cite{Ferrari2021CompilerDF, Diadamo2021DistributedQC}. 
This is mainly due to the dirty side-effect of TP-Comm.
We would need another invocation of
TP-Comm %
to release the occupation of the communication qubit (e.g., $q_0'$ in Figure~\ref{fig:qucomm}(b)), which would be later used for other quantum communications. As a result, two EPR pairs are actually required to implement a single remote CX gate by TP-Comm, with one pair for handling the dirty side-effect.

\begin{figure}[h!] \small
    \centering
    \includegraphics[height=0.175\textwidth]{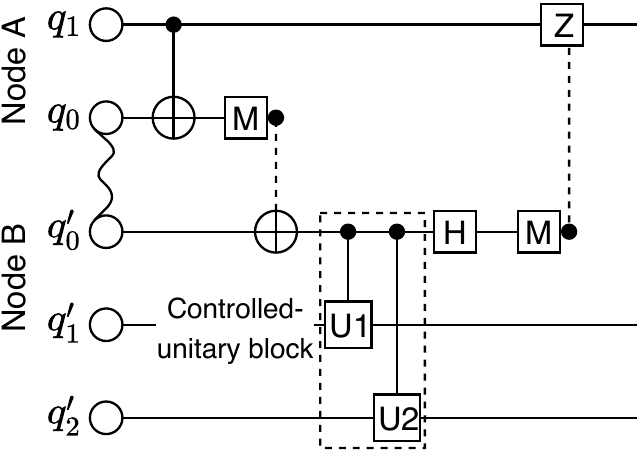}\hfill{} \includegraphics[height=0.175\textwidth]{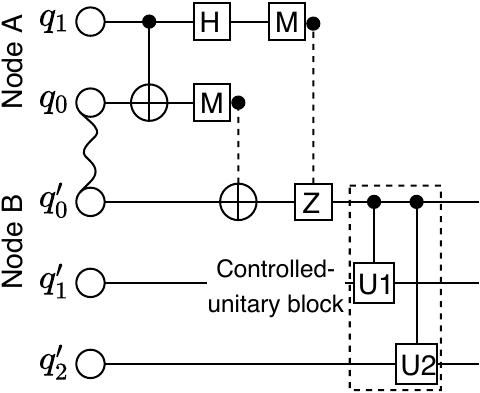} \\
    \hphantom{}\hspace{30pt}(a)\hspace{100pt}(b)
    \caption{The optimized implementation of the controlled-unitary block $C-U1-U2$. (a) The Cat-Comm version. (b) The TP-Comm version. }
    \label{fig:qucomm-cu}
\end{figure}

In Figure~\ref{fig:qucomm}, we only show how to implement one individual CX gate. To implement complex remote interactions between quantum nodes,  one simple strategy is to first decompose the remote interaction into several remote CX gates and implement each remote CX gate as in Figure~\ref{fig:qucomm}. However, this strategy may incur heavy communication costs.
Prior work~\cite{cat_entangle} spots a more efficient way to implement a controlled-unitary block between two quantum nodes. Figure~\ref{fig:qucomm-cu} provides the optimized implementation of the controlled-block $C-U1-U2$, where $U1$ and $U2$ are some unitary quantum operations. The implementation in Figure~\ref{fig:qucomm-cu} only requires one EPR pair, fewer than implementing each remote two-qubit gate independently. %

Besides the controlled-unitary block, we discover that plenty of quantum communications in distributed quantum programs can be transformed into a group of remote interactions between one qubit and one quantum node. We name such a group of remote interactions  \textit{burst communication}. Different from the single CX case, Cat-Comm and TP-Comm each has its own advantage for burst communication of various patterns.
Unfortunately, existing DQC compilers~\cite{ILP, time-slice} either do not take advantage of the burst communication or only consider the basic controlled-unitary case~\cite{Diadamo2021DistributedQC}.

In later sections, we would use \textit{one remote EPR pair} and \textit{one remote communication} interchangeably, because for either Cat-Comm or TP-Comm, one invocation just requires one remote EPR pair.

%% file: 05_algorithm.tex
\section{Problem and Motivation}

In this section, we first introduce the communication problem in distributed quantum programs and then identify the optimization opportunities by considering burst communication. 

For the rest of the discussions, we assume quantum communication can be established between any two quantum nodes, a typical assumption in data-center distributed computing~\cite{van2002distributed}. We also assume that each quantum node  has only two communication qubits, which is realistic for near-term DQC~\cite{Ferrari2021CompilerDF}.

\begin{figure}[tbp]
    \centering
    \includegraphics[width=0.485\textwidth]{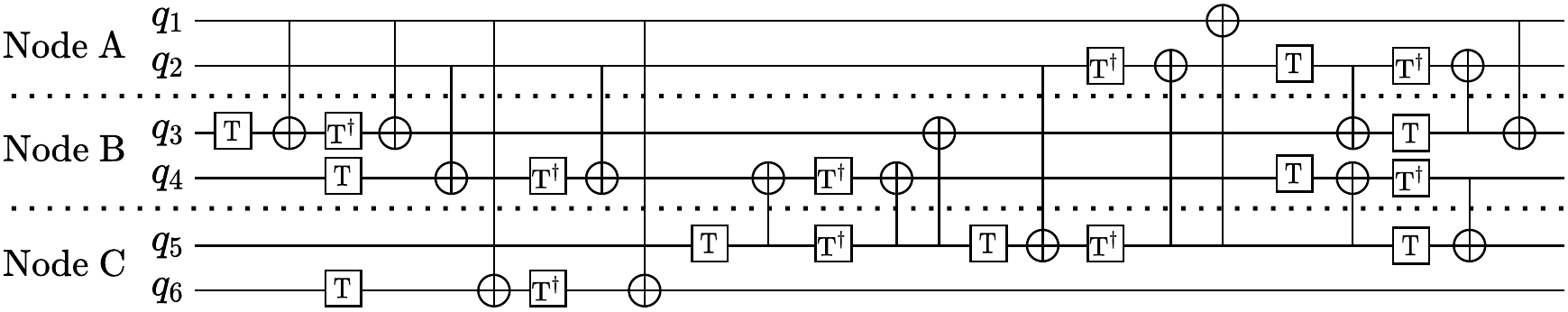} 
        \caption{Program snippet extracted from quantum arithmetic circuits~\cite{revlib}. %
        }
    \label{fig:example}
\end{figure}

\subsection{Communication Problem}

The example distributed program in Figure~\ref{fig:example} is modified from quantum arithmetic circuits~\cite{revlib}.
This program contains many remote CX gates whose control qubit and target qubit reside in different quantum nodes, e.g., $CX\, q_1,q_3$. Remote CX gates are inevitable in DQC especially when the program's qubit number is substantially larger than each quantum node's. To make the distributed program executable, we should transfer
the states of qubits in remote CX gates to make them locally executable temporarily. The state transfer involves remote communication between quantum nodes, which can be accomplished by Cat-Comm or TP-Comm.
Due to the noisy nature of quantum communication,
remote operations are far more error-prone than local quantum gates. The long runtime of quantum communication would also lead to the decoherence of quantum states. As a result, to produce high fidelity outcome, 
we hope the number of remote communication to be as small as possible, so is the latency induced. %

As indicated in Section~\ref{sect: bg}, one remote CX gate requires at least one remote communication.
While there is little room for optimizing the communication cost of one remote CX gate, there is a large optimization space when considering burst communication, which involves a group of remote CX gates. For example, we can execute the first two CX gates on $q_1, q_3$ in Figure~\ref{fig:example} collectively, with only one communication by using the circuit in Figure~\ref{fig:qucomm-cu}(a). From the perspective of information theory, burst communication is more informative than the communication with only one remote CX.
The overall communication cost and latency would be considerably lowered if handling all remote CX gates  in this burst manner.

Fortunately, as we see in the next section, burst communication is prevalent in diverse distributed quantum programs.

\subsection{Burst Communication in DQC}\label{sect:burstcomm}

Aside from the arithmetic program shown in Figure~\ref{fig:example}, we also see  burst communication in a variety of quantum programs. %
As examples, 
we examine the burst communication of the Quantum Fourier Transform  (QFT) program~\cite{nielsen2002quantum} and the Quantum Approximate Optimization Algorithm (QAOA)~\cite{Farhi2014AQA} by hand. These two represent different categories of quantum programs: QAOA is one of the most important applications in near-term quantum computing whereas QFT is the building block circuit of quantum algorithms.

We first give a formal definition of the burst communication in DQC. In this paper, we refer to a group of continuous remote two-qubit gates between one qubit $q$ and one node as \textit{burst communication}. For two remote two-qubit gates $g_1$ and $g_2$, the continuity of these two gates means there are no other remote gates between $g_1$ and $g_2$.

To characterize the burst communication of a distributed program $dprog$, for a remote  gate $g$ in $dprog$, we define function $\epsilon(g)$ to be the largest burst communication block %
that contains $g$. The gate order of $dprog$ may affect the burst communication block found. $\epsilon(g)$ is defined to be the largest over all functional-equivalent gate order of  $dprog$.  
We then define $len(\epsilon(g))$ to be the number of remote CX gates in $\epsilon(g)$ if compiled to the CX+U3 basis~\cite{Qiskit}. 
Finally, we are ready to define the inverse-burst distribution as follows: 
\begin{align}
    P(x) = \dfrac{\vert\{ g \vert len(\epsilon(g)) < x\} \vert}{\# g}.
\end{align}
A lower $P(x)$ suggests more burst communication.

\begin{figure}[tbp]
    \centering
    \includegraphics[width=0.47\textwidth]{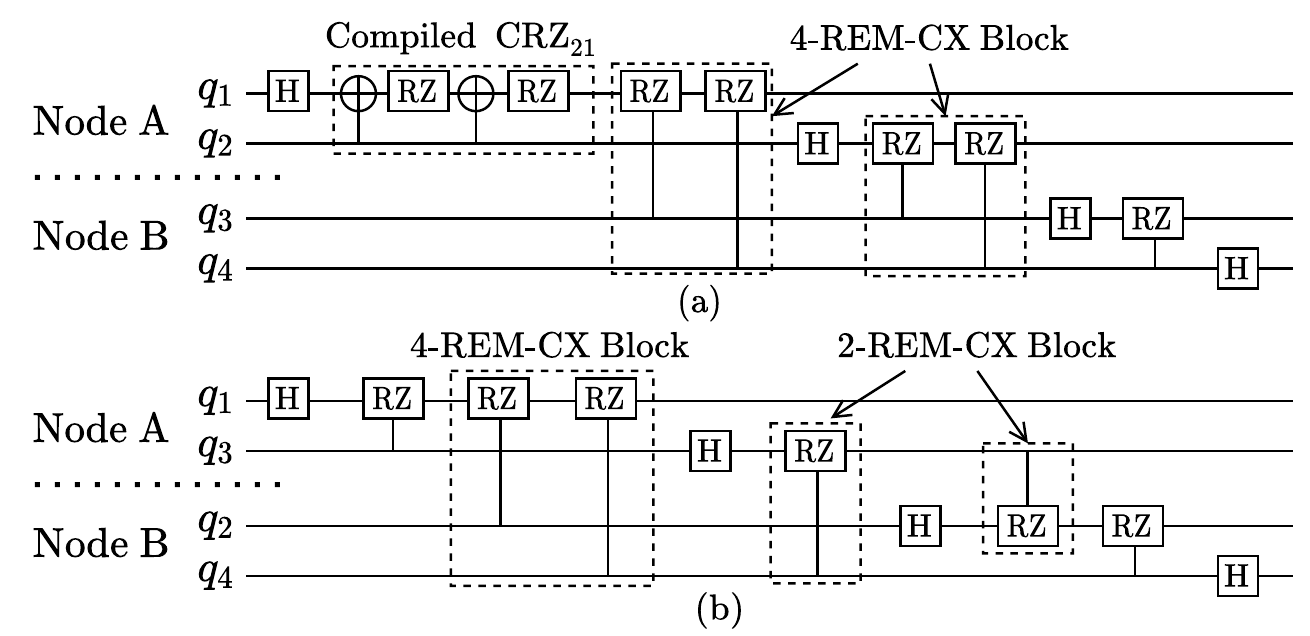}
        \caption{(a) QFT program with two nodes and two qubits per node. (b) The layout for the maximal $P_4$. 
        Parameters omitted for simplicity. 
        For demonstration, we do not combine $\text{CRZ}_{43}$ and $\text{CRZ}_{32}$ to form a 4-REM-CX block.}
    \label{fig:qftexam}
\end{figure}

We begin by examining the QFT program using the aforementioned definition.
We assume the total qubit number is $n$, the quantum node number is $k$, and qubits are evenly distributed across all nodes, with $t = \frac{n}{k}$ qubits per node.  Figure~\ref{fig:qftexam} shows the QFT program with $k=2$ and $t=2$. For the QFT program, as shown in Figure~\ref{fig:qftexam}, each $q_i$ is controlled by all qubits $q_j$ (through the CRZ gate) that satisfies $j > i$~\cite{nielsen2002quantum}. First, we have $P(2) = 0$
because each CRZ gate in QFT is compiled into two CX gates, as illustrated in Figure~\ref{fig:qftexam}(a). Now, we consider $P(4)$. For the $i$-th qubit satisfies $i \le n-k$, the number of $j$ s.t. $\epsilon(\text{CRZ}_{ji}) < 4$ is at most $\lfloor \frac{i-1}{t-1} \rfloor$ because for one node, if at least two of its qubits have subscripts $> i$, 
this node would have at least two qubits being interacted by qubit $i$. Since CRZ gates are commutable with each other, we could form a communication block with at least 4 CX gates. On the other hand, if $i > n-k$, then the $i$-th qubit is at most interacted with $n-i$ qubits, thus the number of $j$ s.t. $\epsilon(\text{CRZ}_{ji}) < 4$ is at most $n-i$.
Therefore, we have 
$$P(4) \le \frac{\sum_{i=1}^{n-k} \lfloor\frac{i-1}{t-1}\rfloor + \sum_{i=n-k+1}^n (n-k)}{\sum_{i=1}^n (n-i) - k\sum_{l=1}^t (t-l) } = \frac{1}{t}.$$
This indicates there are $1-P(4)=1-\frac{1}{t}$ remote gates within a communication block that possesses more than 4 CX gates. Generally, we can prove that $P(2m) \le \frac{m-1}{t}$. This upper bound is quite promising when $t$ is large and it is actually loose. For Figure~\ref{fig:qftexam}(b) which corresponds to the upper bound of $P(4)$, there may be $\frac{1}{t}$ of remote CRZ gates, i.e., $\text{CRZ}_{43}$ and $\text{CRZ}_{32}$ not in a block with 4 remote CX gates at the first glance. But we can actually combine $\text{CRZ}_{43}$ and $\text{CRZ}_{32}$ to form a 4-REM-CX block since there are no other remote gates between them. This indicates that QFT has more abundant burst communication than the upper bound suggests.

\begin{figure}[tbp]
    \centering
    \includegraphics[width=0.47\textwidth]{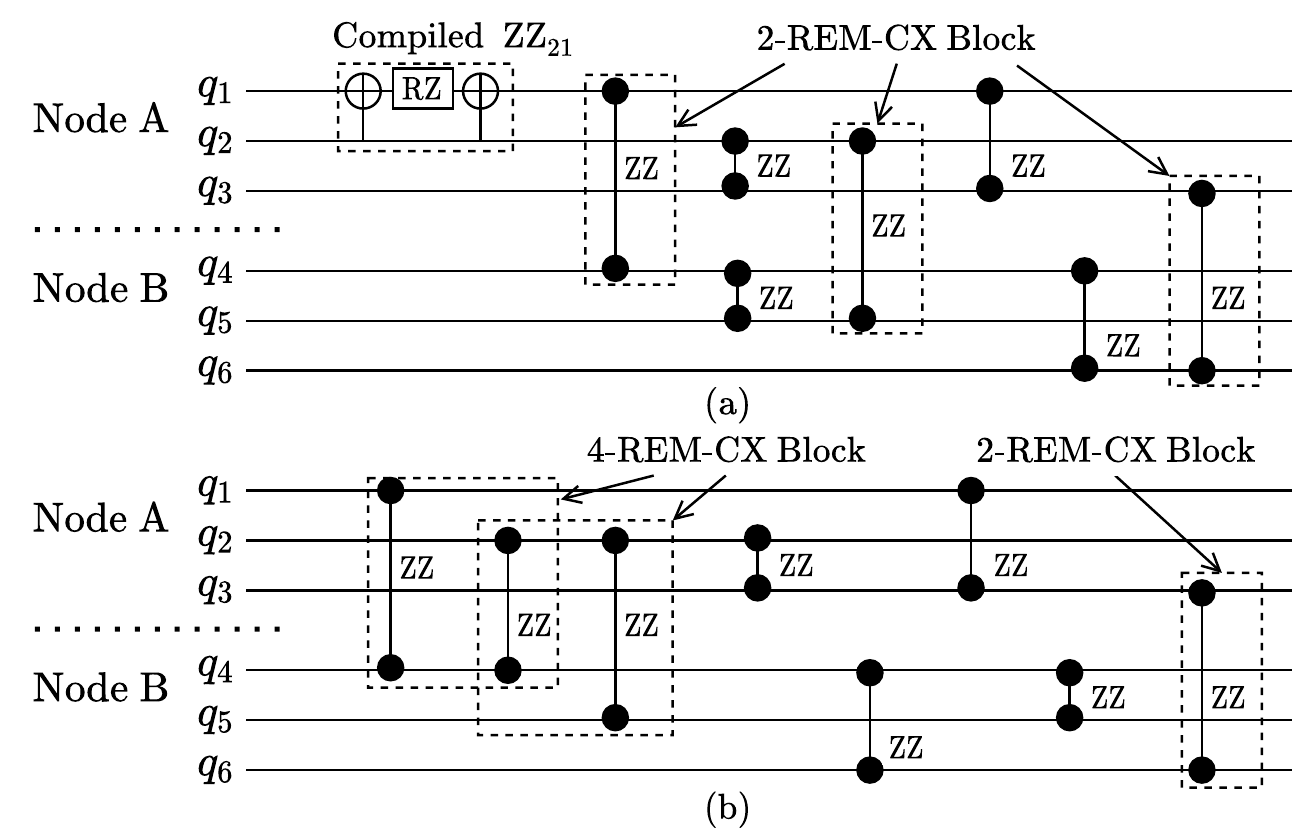}
        \caption{QAOA program with two nodes and three qubits per node. Parameters omitted for simplicity. (a) inter-node communication number $r=3$. (b) $r=4$.}
    \label{fig:qaoaexam}
\end{figure}

Similarly, for the QAOA program, we assume $k$ nodes and $t$ qubits per node. We also suppose  $r$ remote ZZ interactions between any two nodes. Figure~\ref{fig:qaoaexam} shows the QAOA program with $k=2$ and $t=3$. Likewise, $P(2) = 0$ since each ZZ interaction is compiled into two CX gates, as shown in Figure~\ref{fig:qaoaexam}(a). For every two nodes, the qubit layout to minimize $len(\epsilon(ZZ))$ for each ZZ interaction is to make every two ZZ interactions have no shared qubits, i.e., not adjacent. However, this layout at most accommodates $t$ ZZ interactions. If $r > t$, the number of ZZ interactions s.t. $len(\epsilon(ZZ)) < 4$ is at most $t-2(r\ mod\ t)$ by examining the gate adjacency. Thus, $P(4) \le \frac{t-2(r\ mod\ t)}{r}$. For example in Figure~\ref{fig:qaoaexam}(b), only $\frac{t-2(r\ mod\ t)}{r} = \frac{1}{4}$ of remote ZZ interactions are not in a 4-REM-CX block.
Generally, if $r > st$ for some integer $s$, $P(2(s+1)) \le \frac{t-2(r\ mod\ t)}{r} < \frac{1}{s}$. This study reveals that burst communication is broadly available in the QAOA program.

We could derive a similar analysis for other programs.
Further numerical evidence for the richness of burst communication in various programs is shown in Figure~\ref{fig:burst-stat}. The next step is to  figure out how to utilize the abundant burst communication in distributed programs to optimize the communication overhead, as discussed in the next section.

\subsection{Optimization Opportunities}

To exploit burst communication in distributed quantum programs, we need to answer three key questions:

\paragraph{How to unveil the burst communication?} The burst communication is high-level program information and cannot be deduced simply from the low-level circuit language, especially when
the remote interactions between multiple nodes are all mixed together. For example in Figure~\ref{fig:example}, gate $CX\, q_2;q_4$ between node A and node B is followed by $CX\, q_1;q_6$, which is the interaction between node A and node C. To maximize the benefits of burst communication, we need to discover groups of remote gates in disordered quantum circuits.

\paragraph{How to select the best communication scheme?} %
Burst communication comes in various forms. Cat-Comm may not always be better than TP-Comm for burst communication, unlike the single CX case. For example in Figure~\ref{fig:example}, if we use Cat-Comm to implement the last three remote CX gates between $q_3$ and node A, three EPR pairs are needed. However, with TP-Comm to teleport $q_3$ to node A, at most two EPR pairs are needed.
Thus, to reduce the communication cost, we should examine the pattern of burst communication and choose the communication scheme wisely.

\paragraph{How to schedule burst communication?} Finally, we need to schedule the execution of burst communication blocks. If we arrange all burst communication in a sequential way, the large time overhead would impose non-negligible decoherence errors on quantum states. 
As a result, we should maximize the parallelism in burst communication to generate high-fidelity output. To achieve this goal,
we must first identify the relationships between communication blocks and then reduce the gaps caused by them adaptively. %

\section{{\frameworkNameSpace} Framework}

In this section, we first give an overview of the {\frameworkNameSpace} framework and then introduce each component in detail.

\subsection{Design overview}

We propose the \textit{\frameworkName} framework as shown in Figure~\ref{fig:designoverview}. 
{\frameworkName} focuses on the communication optimization of distributed quantum programs and serves as the back-end of front compiling flows like mapping qubits to quantum nodes. We would adopt existing technologies for these front compiling stages, as we would see in Section~\ref{sect:eval}.

To optimize the communication overhead in distributed programs, 
{\frameworkName} comes with three stages to utilize the burst communication. First, it aggregates remote two-qubit gates by gate commutation. Gate commutation is quite common in quantum programs~\cite{Nam2017AutomatedOO}. Commutable gates, on the one hand, may be ordered arbitrarily and hide the burst communication. On the other hand, we could also utilize gate commutation to uncover burst communication blocks. In this stage, a pre-processing step is used to identify burst communication, and a linear merge step is employed to combine isolated burst communication blocks.

Second, it assigns an optimal communication scheme for each burst communication. We observe that the pattern of burst communication impacts the efficiency of communication schemes. Cat-Comm is less expensive for some patterns, while TP-Comm may be more cost-effective for others. It is thus important to examine the communication patterns and consider both Cat-Comm and TP-Comm for hybrid communication, rather than focusing on one scheme.

Third, it performs a block-level schedule of burst communication. 
It is possible to run communication blocks with shared nodes or qubits concurrently or shorten the quantum state transfer path across quantum nodes for specified communication patterns. Combined with these optimizations,
a greedy schedule is effective for burst communication blocks.

\input{img/rule}

\begin{algorithm}[h]
\footnotesize
\SetAlgoLined
\KwIn{An array of communication blocks $blk\_list$}
\KwOut{Merged communication blocks $blk\_list\_merge$}
$blk\_list\_merge = [\,]$\;
$blk = blk\_list[0]$ \;
\While{there are blocks in $blk\_list$ not visited}{
$non\_commute\_gates = []$\;
\For{$blk\_next$ in unvisited blocks of $blk\_list$}{
    \tcp{Attempt merge $blk$ to $blk\_next$}
    \For{$gate$ between $blk$ and $blk\_next$}{
        \uIf{$gate$ is single-qubit and not commutes with $blk$}{
            $non\_commute\_gates.append(gate)$\;
        }
        \uIf{$gate$ is two-qubit}{
            check if $gate$ is commutable with $non\_commute\_gates$ and $blk$\;
            \uIf{not commutable}{
                \uIf{$gate$ is in-node two-qubit}{
                    $non\_commute\_gates.append(gate)$;
                } \uElse{
                    break\;
                }
            }
        }
    }
    $blk = $merge $blk$, $non\_commute\_gates$ and $blk\_next$\;
}
\uIf{the above merge failed}{
    Try to merge $blk\_next$ to $blk$ similarly\;
    \uIf{succeeds}{
        $blk = $merge $blk$, $non\_commute\_gates$ and $blk\_next$\;
    } \uElse{
        $blk = blk\_next$\;
    }
}
}
output the merged blocks and adjust the order of commutable gates\;
\caption{Linear merge procedure}
\label{alg:linear-merge}
\end{algorithm}

\begin{figure}[tbp]
    \centering
    \includegraphics[width=0.485\textwidth]{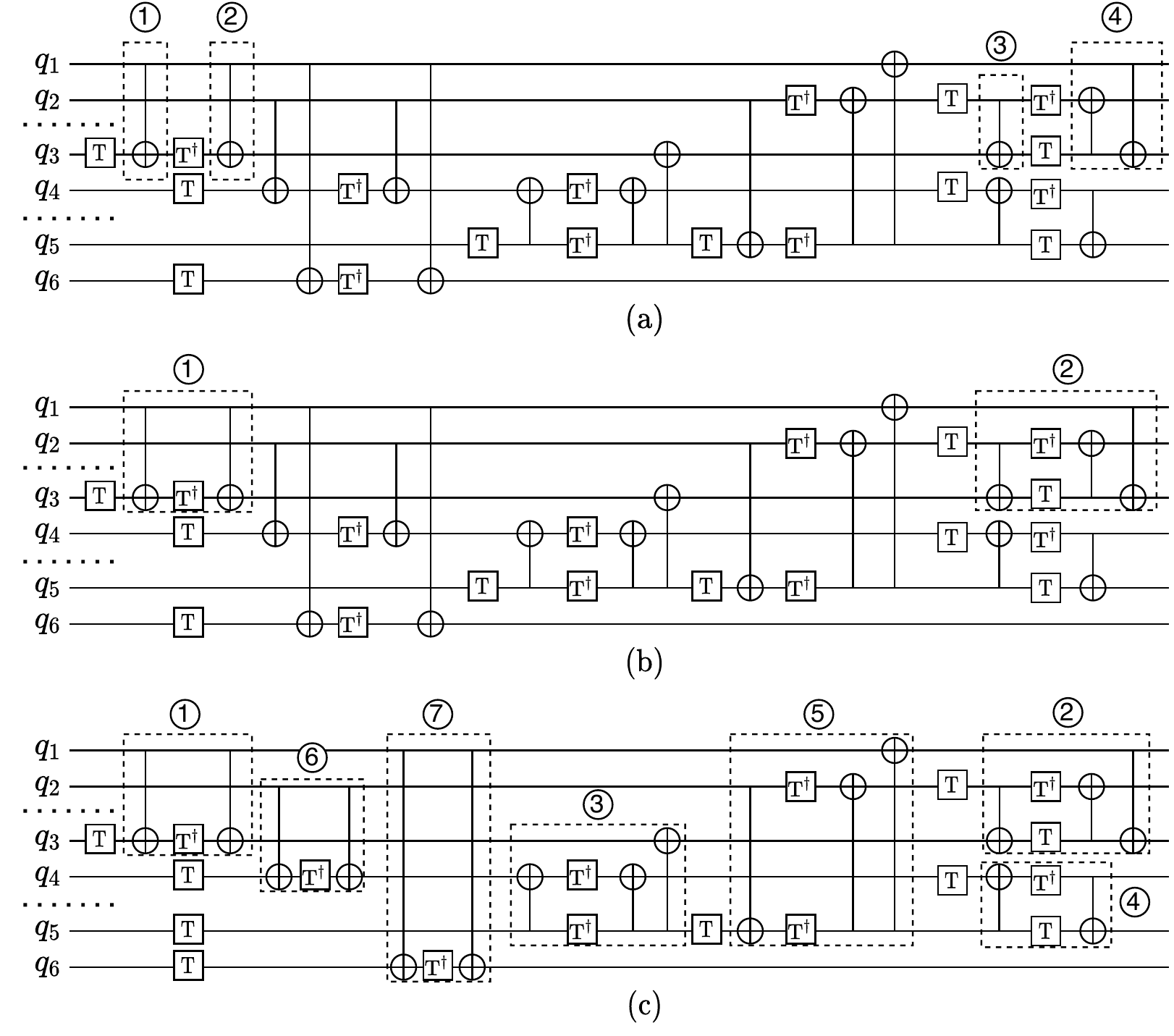}
    \caption{Communication aggregation for the example program in Figure~\ref{fig:example}. (a) Preprocessing. (b) Linear merge. (c) Iterative refinement.}
    \label{fig:commagg}
\end{figure}

\subsection{Communication Aggregating}

Burst communication is prevalent in distributed programs, but may not be immediately available due to two factors: CX gates may be scattered across the program, and whether CX gates are remote depends on the qubit mapping to quantum nodes.
To uncover hidden burst communications, we need to rewrite the circuit and aggregate remote CX gates.

Figure~\ref{fig:xrule} summarizes the X-rotation-centered rewriting rules used for gate commutation. Rules for other rotation axes can be obtained by similar transformation.
Below are the main steps to aggregate remote gates based on these rules. 

\paragraph{Preprocessing:} The first step is to identify the qubit-node pair of burst communication.
We start with the qubit-node pair associated with the most remote gates as it would likely lead to a large burst communication block. For example in Figure~\ref{fig:example}, the chosen qubit-node pair is ($q_3$, node A) as it is associated with 5 remote CX gates. We then search for consecutive remote CX gates related to this qubit-node pair.
This step would result in many isolated communication blocks, for example in Figure~\ref{fig:commagg}(a), we obtain four small blocks.

\paragraph{Linear merge:} The next step is to merge isolated small communication blocks obtained in the preprocessing.
As illustrated in Algorithm~\ref{alg:linear-merge}, we merge related communication blocks in a linear and greedy manner.
For communication blocks \circlenum{1}, \circlenum{2}, \circlenum{3}, \circlenum{4} in Figure~\ref{fig:commagg}(a), 
we can easily merge block \circlenum{1} and \circlenum{2} since only single-qubit gates exist between those two blocks. However, we can not merge block \circlenum{2} and block \circlenum{3} because gate $CX\, q_5,q_3$ is  commutable with neither block \circlenum{2} nor block \circlenum{3}. Finally, as shown in Figure~\ref{fig:commagg}(b), we obtain two larger communication blocks.

\paragraph{Iterative refinement:} Then we merge communication blocks of other qubit-node pairs in descending order of their number of remote gates until no improvement is made. The final result of communication aggregation is shown in Figure~\ref{fig:commagg}(c).

\subsection{Communication Assignment}\label{sect:blocking}

With burst communication blocks, the next optimization is to find the best way to execute them. 
We address this problem 
by first examining the pros and cons of Cat-Comm and TP-Comm, and then assigning the optimal communication scheme based on the pattern analysis of burst communication blocks. Since we assume only two communication qubits in each quantum node, the communication patterns discussed here center on interactions between one qubit and one node. Extending burst communication to the node-to-node situation is promising when communication qubits are plentiful. We leave it for future work.

\paragraph{Cat-Comm vs. TP-Comm:} 
Suppose we have a burst communication block between a qubit $q_1$ in node A and several qubits in node B, with a total of $n$ remote CX gates in the block. If the block can be executed by a single call to Cat-Comm, the savings on EPR pairs would be up to $n$ times, compared to executing each remote CX gate individually. However, as discussed below, not all communication blocks can be cheaply executed via Cat-Comm. Compared to Cat-Comm, the savings on ERP pairs with TP-Comm is at most $\frac{n}{2}$ times as TP-Comm requires two EPR pairs to execute any burst communication block: one to teleport $q_1$ to node B, the other to release the occupancy of $q_1$ on the communication qubit in node B. For simplicity, we use the other EPR pair to teleport $q_1$ back to node A. 
We postpone to %
Section~\ref{sect:schedule} to handle the case that teleporting $q_1$ to some other node is better than moving back.
Overall, Cat-Comm provides higher ERP pair savings for specific burst communication blocks, while TP-Comm can handle an arbitrary communication block with up to two EPR pairs. %

\begin{figure}[tbp]
    \centering
    \includegraphics[width=0.45\textwidth]{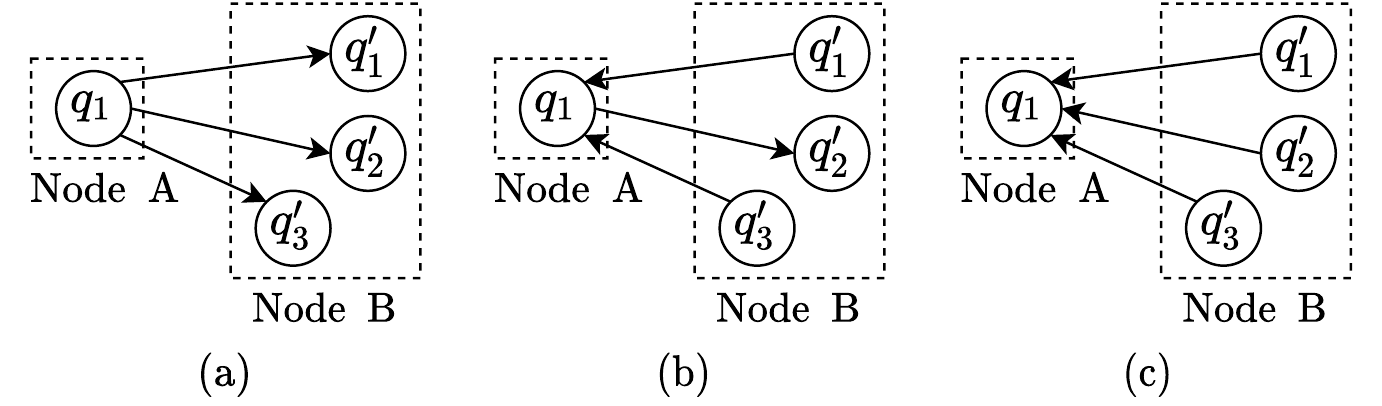}
    \caption{Two primitive communication patterns (a)(b) and the variant (c).} %
    \label{fig:pattern1}
\end{figure}

\begin{figure}[tbp]
    \centering
    \includegraphics[width=0.45\textwidth]{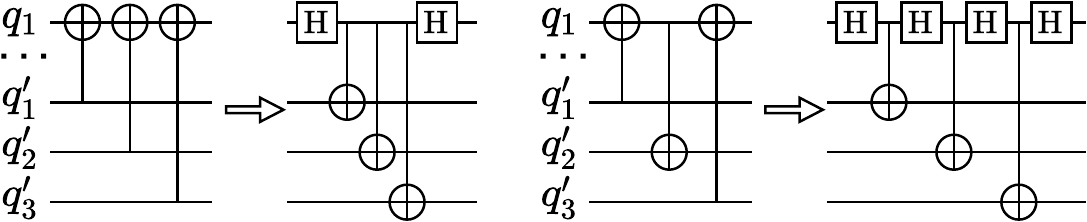}\\
    \hphantom{} \hspace{0pt}(a) \hspace{106pt}(b)
    \caption{The transformation between communication patterns by using Hadamard gates.}
    \label{fig:patterntransform}
\end{figure}

\begin{figure*}[t]
    \centering
    \includegraphics[width=0.98\textwidth]{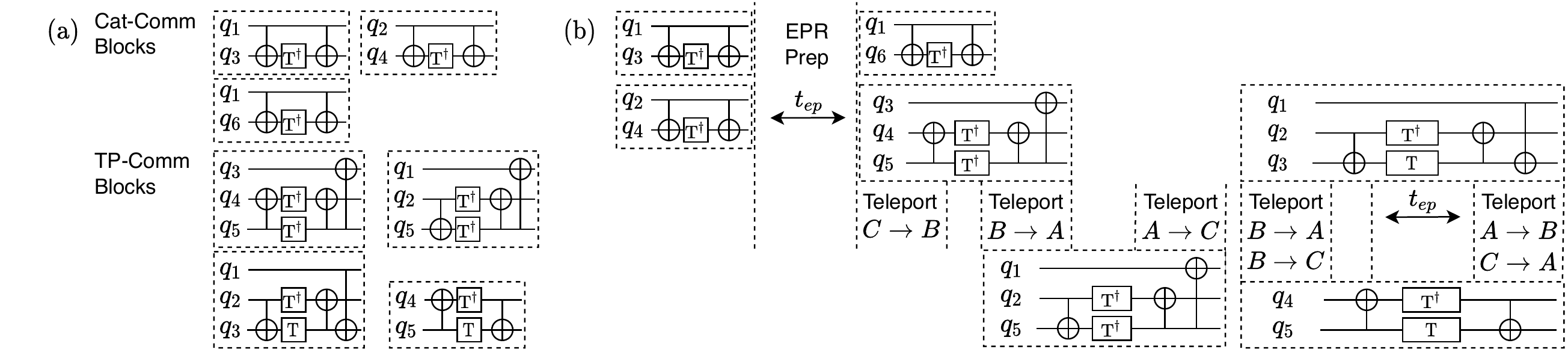}
    \caption{(a) The result of the communication assignment pass. (b) The result of the communication schedule pass.}
    \label{fig:commassignsched}
\end{figure*}

\paragraph{Pattern analysis:}
Figure~\ref{fig:pattern1}(a)(b) shows two primitive  patterns for qubit-to-node burst communication. 
For the unidirectional communication pattern in Figure~\ref{fig:pattern1}(a) where one qubit (i.e., $q_1$) always serves as the control qubit, the communication block can be implemented by Cat-Comm with only one EPR pair  if
no single-qubit gate on the control qubit separates two-qubit gates~\cite{cat_entangle}. For example, one call of Cat-Comm can handle the gate sequence $CX\,q_1,q_1';\,CX\,q_1,q_2'$ , but cannot address  $CX\,q_1,q_1';\,H\,q_1;\,CX\,q_1,q_2'$ due to the middle H gate. 
To optimize this communication pattern with Cat-Comm, we should remove single-qubit gates on the control qubit. When they are not removable, we resort to TP-Comm.  %

A varied unidirectional pattern in which $q_1$ always serves as the target qubit, as shown in Figure~\ref{fig:pattern1}(c), also occurs frequently in distributed quantum programs. This pattern can be transformed into the pattern in Figure~\ref{fig:pattern1}(a) by applying a series of Hadamard gates, as shown in Figure~\ref{fig:patterntransform}(a).

In contrast to unidirectional patterns, Figure~\ref{fig:pattern1}(b) shows a bidirectional pattern in which $q_1$ serves as both control qubit and target qubit. A block in this pattern cannot be executed by a single call of Cat-Comm as Cat-Comm cannot transfer the state of target qubits. Even if we transform it to the unidirectional pattern in Figure~\ref{fig:patterntransform}(b) with Hadamard gates,  single-qubit gates on the control qubit still prevent a cheap implementation by Cat-Comm. In fact, for the block in Figure~\ref{fig:patterntransform}(b), TP-Comm is more efficient as it only requires two EPR pairs, while Cat-Comm requires three EPR pairs.

To summarize, for unidirectional patterns in Figure~\ref{fig:pattern1}(a)(c), we will try Cat-Comm first, while for the bidirectional pattern in Figure~\ref{fig:pattern1}(b), TP-Comm is preferred.

\paragraph{Communication assignment:}
Now, we are ready to assign an optimal communication scheme, either Cat-Comm or TP-Comm, to each burst communication block. 
Considering Figure~\ref{fig:commagg}(c) as an example, we assign Cat-Comm to unidirectional blocks \circlenum{1}, \circlenum{6} and \circlenum{7}, and assign TP-Comm to bidirectional blocks \circlenum{2}, \circlenum{4} and \circlenum{5}. For \circlenum{3}, although being unidirectional, it cannot be executed by one call of Cat-Comm as there is a $T^{\dagger}$ gate on the control qubit between two CX gates. Since executing it with either Cat-Comm or TP-Comm  requires two EPR pairs, we set the TP-Comm assignment as default. The finalized assignment is shown in Figure~\ref{fig:commassignsched}(a).

\subsection{Communication Scheduling}\label{sect:schedule}

After optimizing the count of remote communications, we then schedule the execution of burst communication blocks to
reduce the total execution time of the distributed  program and reduce the impact of decoherence. Based on the quantitative data shown in Table~\ref{tab:quantdata}, the preparation of remote EPR pairs is the most time-consuming one among various operations and hence should be carefully optimized to hide its latency. While the quantitative data may vary across quantum devices, the schedule design in this section should be also effective.

\begin{table}[htbp]
\centering
\small
\renewcommand*{\arraystretch}{1}
\begin{tabular}{|l|c|l|}
\hline
Operation & Variable Name & Latency \\ \hline
Single-qubit gates & $t_{1q}$ & $\sim$ 0.1 CX \\ \hline
CX and CZ gates & $t_{2q}$ & 1 CX \\ \hline
Measure & $t_{ms}$ & 5 CX \\ \hline
EPR preparation %
& $t_{ep}$ & $\sim$ 12 CX \\ \hline
One-bit classical comm & $t_{cb}$ & $\sim$ 1 CX \\ \hline
\end{tabular}
\caption{The quantitative latency data of operations in distributed quantum programs, extracted from~\cite{isailovic2006interconnection, correa2018ultra}. All latencies are normalized to CX counts.} 
\label{tab:quantdata}
\end{table}

The designs here aim to maximize block-level parallelism and shorten the latency of sequential execution by fusion.

\paragraph{More block-level parallelism:}
The essence of scheduling is to maximize the parallelism in a circuit. 
For burst communication blocks without nodes or qubits in common, they can be concurrently executed in nature. For blocks with shared nodes or qubits, their parallelism is limited by their commutability, as well as the communication resources each node holds. With the constraint that each node can establish only two communications in parallel, there is little room for lazy operations, and we adopt a greedy strategy to execute commutable blocks, i.e., execute as many blocks as possible simultaneously, as soon as EPR pairs are prepared. 

\begin{figure}[h]
    \centering
    \includegraphics[width=0.36\textwidth]{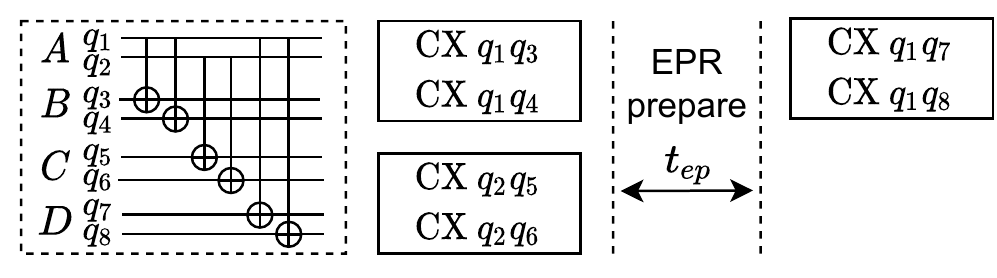}\vspace{-8pt}
    \caption{The schedule optimization for commutable Cat-Comm blocks, with shared qubit or node. %
    }
    \label{fig:pipesched}
\end{figure}

\begin{figure}[h]
    \centering
    \includegraphics[width=0.49\textwidth]{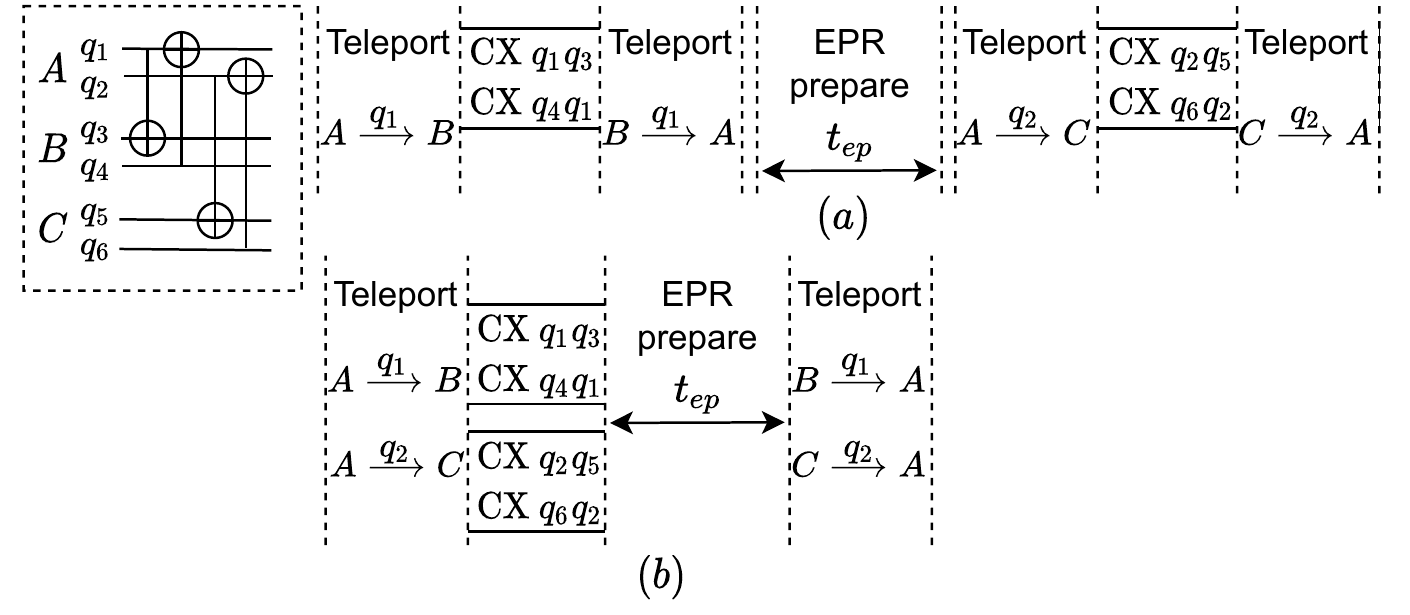}\vspace{-8pt}
    \caption{The schedule optimization for TP-Comm blocks.  Aligned qubit teleportation in (b) is better than the independent qubit teleportation in (a). %
    }
    \label{fig:greedytp}
\end{figure}

For Cat-Comm blocks, we can execute two commutable blocks in parallel at most if they share nodes, as shown in Figure~\ref{fig:pipesched}. For TP-Comm blocks, the situation is complex as each TP-Comm blocks require two EPR pairs. For two commutable TP-blocks, rather than prioritizing the completion of one TP-comm as in Figure~\ref{fig:greedytp}(a), we observe that parallelism can be enabled by communication alignment, as shown in Figure~\ref{fig:greedytp}(b). Compared to Figure~\ref{fig:greedytp}(a), Figure~\ref{fig:greedytp}(b) aligns the qubit teleportation of the two blocks, leading to a latency saving of $t_{block}+2t_{tele}$. This TP-Comm alignment technique can be generalized to the case of $n$ commutable TP-Comm blocks (any two blocks may share common nodes). With TP-Comm alignment, the total latency saving can be up to $(n-1)(t_{block}+2t_{tele})$ (e.g., if those TP-Comm blocks are on nodes $\{A_1, A_2\}, \{A_2, A_3\},\cdots, \{A_n,  A_{n+1}\}$ respectively).

\begin{figure}[h]
    \centering
    \includegraphics[width=0.48\textwidth]{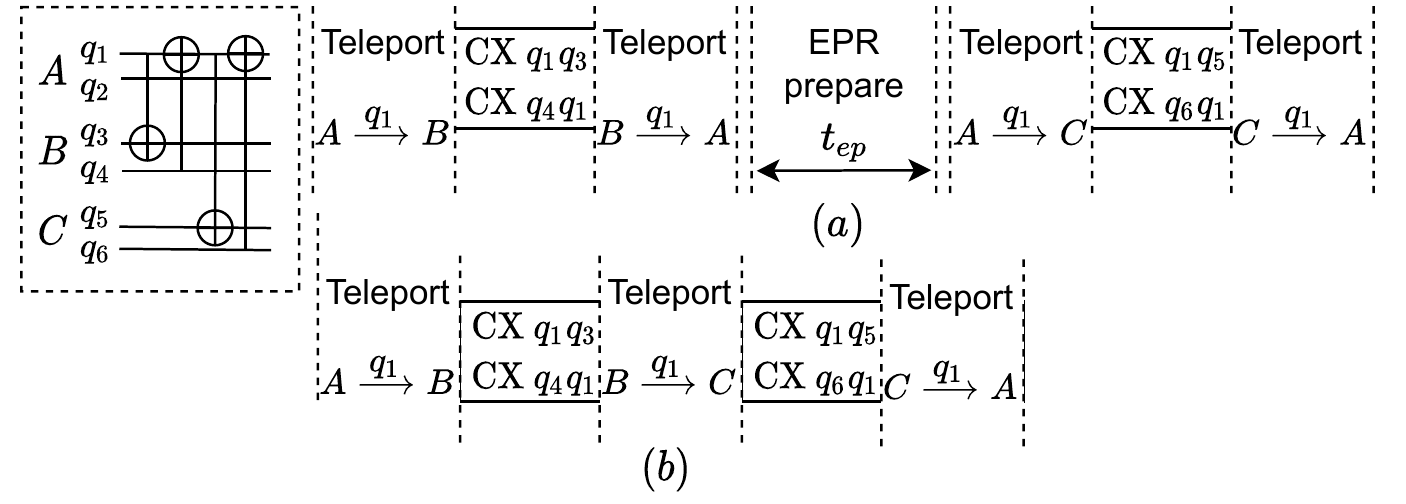}\vspace{-8pt}
    \caption{The schedule optimization for TP-Comm blocks.  Cyclic qubit teleportation in (b) is better than the SWAP-style qubit teleportation in (a). %
    }
    \label{fig:tokenpass}
\end{figure}

\paragraph{Fusion of sequential blocks:}
Sometimes communication blocks have to be executed in sequence. However, if the teleported qubits in TP-Comm blocks are the same, we can optimize their executions by fusing the teleportations, as shown in Figure~\ref{fig:tokenpass}. Figure~\ref{fig:tokenpass}(a) shows a simple schedule where each TP-Comm is executed independently. As each node has only two communication qubits, we need to wait for $t_{ep}$ before executing the next TP-Comm block. In contrast, Figure~\ref{fig:tokenpass}(b) fuses the teleportations between quantum nodes, forming a cycle: $A \to B \to C \to A$. With fusion, the number of teleportations is reduced by one and the overall execution time is reduced by $t_{ep}+t_{tele}$, where $t_{tele}$ is the time to teleport one qubit (about 8 CX time as shown in Figure~\ref{fig:qucomm}(b)). Generally, if we have $n$ TP-Comm blocks with the same teleported qubit, the total number of teleportation would be reduced by $n-1$, and the saving of overall latency would be $(n-1)(t_{ep}+t_{tele})$.

From another view, the fusion also optimizes the token passing problem in classical distributed computing~\cite{van2002distributed}, which also appears in Section~\ref{sect:blocking}, about whether to move the teleported qubit in TP-Comm back or to another node.

With the designs above, the communication schedule pass should apply block-level commutation analysis to unveil the patterns discussed above and then apply corresponding optimizations. We omit the details since this procedure is very similar to the communication aggregation except working at the block level.
With all those optimizations applied, Figure~\ref{fig:commassignsched}(b) shows the optimized communication schedule for the program in Figure~\ref{fig:example}. In total, 2.4x latency saving is achieved compared to executing each remote CX gate independently.

%% file: img/rule.tex
\begin{figure}[tb]
\newcommand{\cfr}{0.42}
\begin{center}
    \begin{adjustbox}{width=\cfr\textwidth}        
        \begin{quantikz}[row sep=0.1cm, column sep=0.3cm, align equals at=1]
            & \gate{X}  & \gate{P}  & \qw 
        \end{quantikz}
        =\begin{quantikz}[row sep=0.1cm, column sep=0.3cm, align equals at=1]
            & \gate{P^\dagger} &  \gate{X}  & \qw
        \end{quantikz}
        \quad
        \begin{quantikz}[row sep=0.1cm, column sep=0.3cm, align equals at=1]
            & \gate{H} & \gate{RX}  & \qw
        \end{quantikz}
        =    
        \begin{quantikz}[row sep=0.1cm, column sep=0.3cm, align equals at=1]
            & \gate{RZ} & \gate{H}  & \qw
        \end{quantikz}
\end{adjustbox}
\end{center} \vspace{-20pt}
\begin{center}  
\begin{adjustbox}{width=\cfr\textwidth}        
    \begin{quantikz}[row sep=0.1cm, column sep=0.2cm, align equals at=1.5]
        & \ctrl{1} & \qw & \qw\\
        & \targ{} & \gate{RX} & \qw
    \end{quantikz}
    \quad=\quad
    \begin{quantikz}[row sep=0.1cm, column sep=0.2cm, align equals at=1.5]
        & \qw  & \ctrl{1} & \qw\\
        & \gate{RX} & \targ{} & \qw
    \end{quantikz}
    \qquad
    \begin{quantikz}[row sep=0.1cm, column sep=0.2cm, align equals at=1.5]
        & \ctrl{1}  & \gate{RZ}  & \qw\\
        & \targ{}  & \qw & \qw
    \end{quantikz}
    \quad=\quad    
    \begin{quantikz}[row sep=0.1cm, column sep=0.2cm, align equals at=1.5]
        & \gate{RZ}  & \ctrl{1} & \qw\\
        & \qw & \targ{} & \qw
    \end{quantikz}
\end{adjustbox}
\end{center}\vspace{-20pt}

\begin{center}
    \begin{adjustbox}{width=\cfr\textwidth}        
        \begin{quantikz}[row sep=0.1cm, column sep=0.2cm, align equals at=2]
            & \ctrl{1} & \qw & \qw\\
            & \targ{} & \targ{} & \qw\\
            & \qw & \ctrl{-1} & \qw
        \end{quantikz}
        \quad=\quad
        \begin{quantikz}[row sep=0.1cm, column sep=0.2cm, align equals at=2]
            & \qw & \ctrl{1} & \qw\\
            & \targ{} & \targ{} & \qw\\
            & \ctrl{-1} & \qw & \qw
        \end{quantikz}
        \qquad
        \begin{quantikz}[row sep=0.1cm, column sep=0.2cm, align equals at=2]
            & \ctrl{1} & \ctrl{2} & \qw\\
            & \targ{} & \qw & \qw\\
            & \qw & \targ{} & \qw
        \end{quantikz}
        \quad=\quad
        \begin{quantikz}[row sep=0.1cm, column sep=0.2cm, align equals at=2]
            & \ctrl{2} & \ctrl{1} & \qw\\
            & \qw & \targ{} & \qw\\
            & \targ{} & \qw & \qw
        \end{quantikz}
\end{adjustbox}
\end{center}\vspace{-20pt}

\begin{center}
    \begin{adjustbox}{width=\cfr\textwidth}        
        \begin{quantikz}[row sep=0.1cm, column sep=0.2cm, align equals at=1.5]
            & \qw & \ctrl{1} & \qw & \qw\\
            & \gate{H} & \targ{} & \gate{H} & \qw
        \end{quantikz}   
        \quad=\quad
        \begin{quantikz}[row sep=0.1cm, column sep=0.2cm, align equals at=1.5]
            & \qw & \ctrl{1} & \qw & \qw \\
            & \ghost{H}\qw & \control{} & \ghost{H}\qw & \qw
        \end{quantikz}
        \qquad
        \begin{quantikz}[row sep=0.1cm, column sep=0.2cm, align equals at=1.5]
            & \gate{X} & \ctrl{1} & \qw\\
            & \ghost{X}\qw & \targ{} & \qw
        \end{quantikz}
        \quad=\quad
        \begin{quantikz}[row sep=0.1cm, column sep=0.2cm, align equals at=1.5]
            & \ctrl{1} & \gate{X} & \qw\\
            & \targ{} & \gate{X} & \qw
        \end{quantikz}
\end{adjustbox}\vspace{-10pt}
\end{center}
    \caption{Phase gates P includes $Z, RZ, S, T$, etc. X-rotation-centered rules for gate commutation. 
    }
    \label{fig:xrule}
\end{figure}

%% file: 07_evaluation.tex
\section{Evaluation}\label{sect:eval}

In this section, we first compare the performance of {\frameworkName} to the baseline method and then evaluate the effect of each optimization pass in {\frameworkName}. We finally perform a sensitivity analysis on {\frameworkName} to study how its performance evolves as the program configuration changes.

\subsection{Experiment Setup}

\paragraph{Metric} The first metric we considered is the number of issued remote communications. Each remote communication would consume one remote EPR pair for both Cat-Comm and TP-Comm. To avoid the ambiguity on the cost of TP-Comm, we say TP-Comm needs two communications (i.e., two EPR pairs) to execute one burst communication block, with one of the communications handling its dirty side-effect.
The number of remote communications models the resource overhead of executing distributed programs and a lower value is favored. 

The second metric is the maximum number of remote two-qubit gates executed through one communication. For TP-Comm blocks, this number is averaged on two communications. We denote this metric by `Peak \# REM CX'.
This metric models the communication throughput of information and a higher value is preferred. 

Finally, we consider two metrics that model the relative performance, in communication cost and program latency respectively, of {\frameworkName} to baselines. The first one is `improv. factor', which is defined to be `total communication \# by baseline/total communication \# by \frameworkName'. The second one is `LAT-DEC factor' and is defined to be `program latency by baseline/program latency by  \frameworkName'. We hope these two metrics to be as large as possible.

\paragraph{Baseline} For the baseline method, we implement the compiler~\cite{Ferrari2021CompilerDF} which only exploits the Cat-Comm scheme for remote CX gates and does not consider burst communication. To reduce the program latency, the baseline adopts a greedy scheduling method, i.e., executing operations as soon as possible.
For both the baseline and {\frameworkName}, we map qubits to distributed quantum nodes by the `Static Overall Extreme Exchange' strategy studied in ~\cite{time-slice}.

\paragraph{Platforms} We perform all experiments on a Ubuntu 18.04 server with a 6-core Intel E5-2603v4 CPU and 32GB RAM. Other software includes Python 3.8.3 and Qiskit 0.18.3~\cite{Qiskit}.

\paragraph{Benchmark programs} We consider two categories of benchmark programs, as shown in Table~\ref{tab:benchmark}. The first category of benchmarks are function-specific, i.e., they focus on implementing specific elementary functions, e.g., arithmetic operations and Fourier transformation. These quantum programs are often used as the building blocks of large quantum applications. The second category of benchmarks are application-driven. These programs usually target at solving real-world problems, e.g., Bernstein-Vazirani (BV) algorithm, Quantum Approximate Optimization algorithm (QAOA), and Unitary Coupled Cluster ansatzes (UCCSD). Specifically, we choose the graph maxcut problem for the QAOA test programs, and for the UCCSD programs, we select molecules $\text{LiH}, \text{BeH}_2$, and $\text{CH}_4$ which correspond to 8, 12, and 16 particles (thus qubits), respectively.
All benchmark programs used in the evaluation are collected from IBM Qiskit~\cite{Qiskit} and RevLib~\cite{revlib}. 

\begin{table}[t]
    \centering
    \resizebox{0.4\textwidth}{!}{
    \renewcommand*{\arraystretch}{1.4}
    \begin{tabular}{|p{1.0cm}|p{1.5cm}|p{1.0cm}|p{1.3cm}|p{1.0cm}|p{1.0cm}|p{1.55cm}|}  \hline
    Type & Name & \# qubit & \# node & \# gate & \# CX & \# REM CX \\ \hline
\multirow{9}{1.0cm}{Build-ing Blocks} & \multirow{3}{1.5cm}{Multi-Controlled Gate (MCTR)} & 100 & 10 & 10640 & 4560 & 1680 \\ \cline{3-7} 
 &  & 200 & 20 & 21840 & 9360 & 3568 \\ \cline{3-7} 
 &  & 300 & 30 & 33040 & 14160 & 5632 \\ \cline{2-7} 
 & \multirow{3}{1.5cm}{Ripple-Carry Adder (RCA)} & 100 & 10 & 1569 & 785 & 220 \\ \cline{3-7} 
 &  & 200 & 20 & 3169 & 1585 & 662 \\ \cline{3-7}
 &  & 300 & 30 & 4769 & 2385 & 820 \\ \cline{2-7} 
 & \multirow{3}{1.5cm}{Quantum Fourier Transform (QFT)} & 100 & 10 & 40100 & 20000 & 9000  \\ \cline{3-7} 
 &  & 200 & 20 & 160200 & 80000 & 38000 \\ \cline{3-7} 
 &  & 300 & 30 & 360300 & 180000 & 87000 \\ \hline
\multirow{9}{1.0cm}{Real World Appli-cations} & \multirow{3}{1.5cm}{Bernstein Vazirani (BV)} & 100 & 10 & 265 & 65 & 56 \\ \cline{3-7} 
 &  & 200 & 20 & 535 & 135 & 126 \\ \cline{3-7} 
 &  & 300 & 30 & 803 & 203 & 194 \\ \cline{2-7} 
 & \multirow{3}{1.5cm}{QAOA} & 100 & 10 & 6000 & 4000 & 3144 \\ \cline{3-7} 
 &  & 200 & 20 & 24000 & 16000 & 14076 \\ \cline{3-7} 
 &  & 300 & 30 & 54000 & 36000 & 32896 \\ \cline{2-7} 
 & \multirow{3}{1.5cm}{UCCSD} & 8 & 4 & 3129 & 1420 & 900  \\ \cline{3-7} 
 &  & 12 & 6 & 40659 & 19142 & 15136 \\ \cline{3-7} 
 &  & 16 & 8 & 129829 & 64956 & 53426 \\ \hline
\end{tabular}
    }
   \caption{Benchmark programs. The qubits are evenly distributed across quantum nodes. The number of remote CX gates (\# REM CX) is computed on the qubit mapping by `Static Overall Extreme Exchange'~\cite{time-slice}.}
    \label{tab:benchmark}
\end{table}

\begin{figure}
    \includegraphics[width=0.23\textwidth]{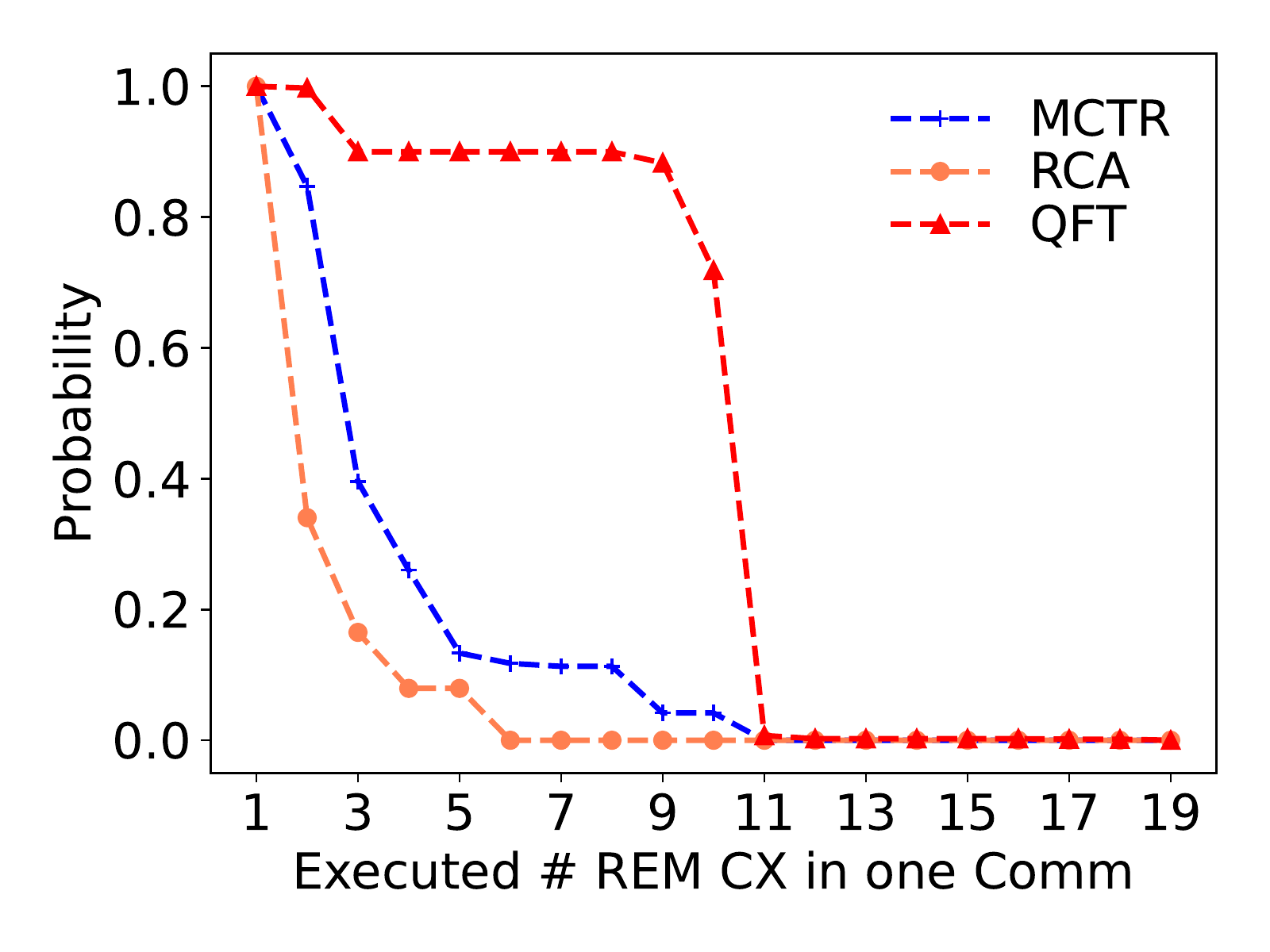}\hfill \includegraphics[width=0.23\textwidth]{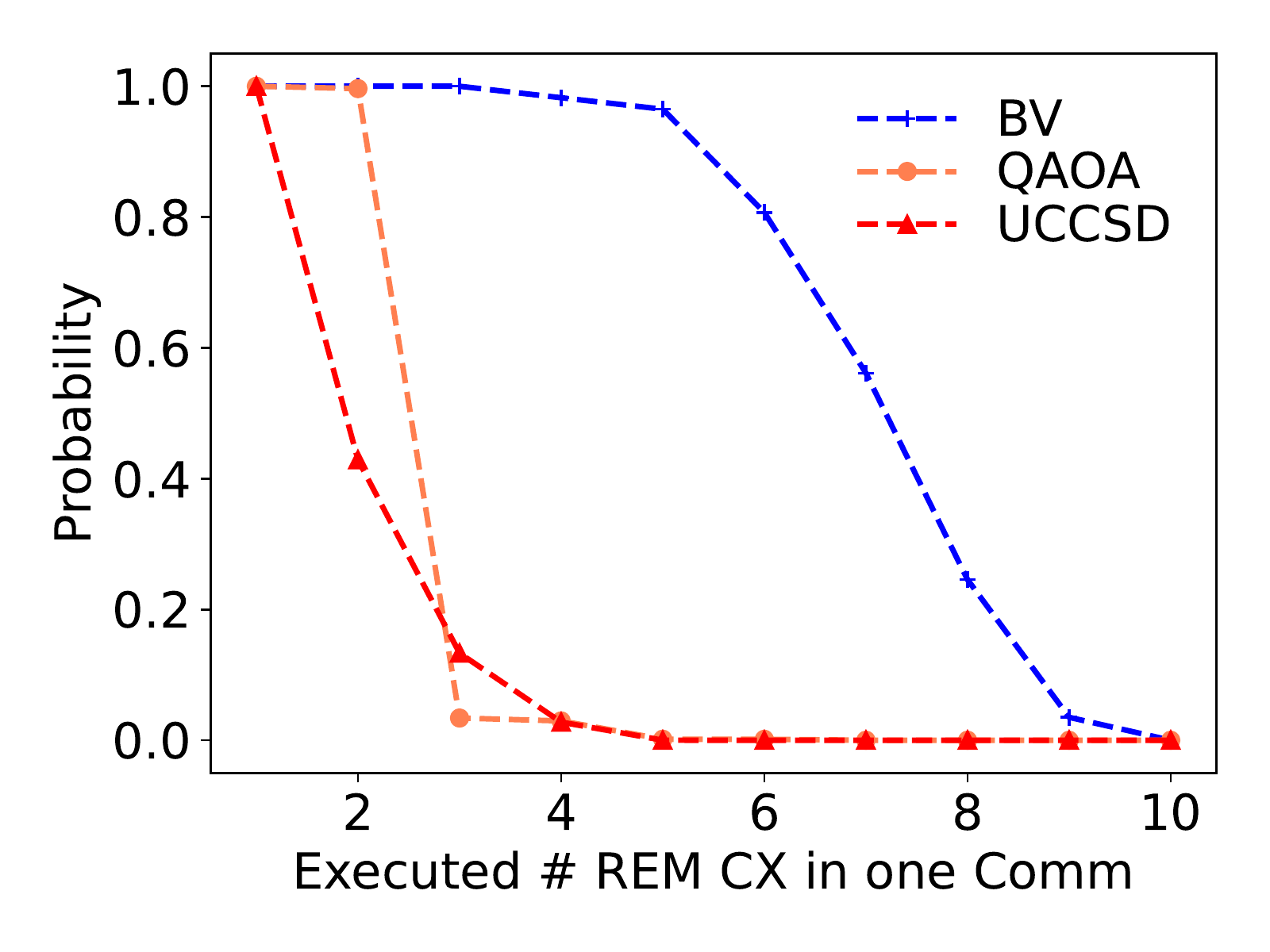} \\
        \hphantom{} \hspace{60pt}(a) \hspace{108pt} (b)
    \caption{Burst communications by \frameworkName: Pr[X] = Pr[one communication carries >= X\hspace{3pt} REM-CXs].}
    \label{fig:burst-stat}
\end{figure}

\subsection{Compared to Baseline}

We evaluate both {\frameworkName} and the baseline method on benchmark programs in Table~\ref{tab:benchmark}. The results of {\frameworkName} and its relative performance to the baseline are shown in Table~\ref{tab:comm-cost}.

\paragraph{Burst communication statistics:} Figure~\ref{fig:burst-stat} shows the distribution of burst communications assembled by {\frameworkName}. This distribution is closely related to the inverse-burst distribution discussed in Section ~\ref{sect:burstcomm} but is easier to compute. We can see that burst communications exist widely in various distributed quantum programs, no matter in building-block circuits (Figure~\ref{fig:burst-stat}(a)) or in real-world applications (Figure~\ref{fig:burst-stat}(b)). Moreover, Figure~\ref{fig:burst-stat} demonstrates the effectiveness of {\frameworkName} in unveiling burst communications. In Figure~\ref{fig:burst-stat}, the communications that each carries $\ge2$ remote CX gates account for 76.8\% of the total remote communications, on average.

\paragraph{Communication cost:} {\frameworkName} achieves significant communication cost reduction on the benchmark programs.
Compared to the baseline method, {\frameworkName} reduces the number of remote communications by a factor of 4.1x on average, up to 9.2x. The peak communication throughput (i.e., `Peak \#\,REM CX') by {\frameworkName} is 8.8x on average and up to 18x of that by the baseline. 
These improvements indicate that {\frameworkName} can efficiently discover and utilize burst communications in distributed quantum programs, transferring more information in each communication than the baseline method. 

The good communication performance of {\frameworkName} comes from two factors: the aggregation of remote CX gates and the hybrid implementation of burst communications by using both Cat-Comm and TP-Comm. We will further elaborate on this point in Section~\ref{sect:optianaly}.

\paragraph{Latency:} {\frameworkName} also achieves significant latency reduction on benchmark programs.
Compared to the baseline method, {\frameworkName} reduces the program execution time by a factor of 3.5x on average, up to 7.1x, as shown in Table~\ref{tab:comm-cost}.
The trend of latency reduction is closely related to the trend of communication cost reduction. This is as expected because {\frameworkName}  keeps the local parallelism in the program when aggregating remote interactions.

\begin{table}[tbp]
\centering
\footnotesize
\resizebox{0.48\textwidth}{!}{
\renewcommand*{\arraystretch}{1.2}
\begin{tabular}{|p{2.cm}|p{1.35cm}|p{1.35cm}|p{1.1cm}|p{1.0cm}|p{1.29cm}|}  \hline
{\vspace{0.2pt}Name} &  {\vspace{0.2pt}Tot Comm} & {\vspace{0.2pt}TP-Comm} & Peak\,\# REM\,CX  & Improv. factor & LAT-DEC factor   \\ \hline
MCTR-100-10 & 533 & 220 & 10 & 3.15 & 3.27 \\\hline
MCTR-200-20 & 972 & 418 & 10 & 3.67 & 3.83 \\\hline
MCTR-300-30 & 2044 & 1112 & 10 & 2.76 & 2.88 \\\hline
RCA-100-10 & 79 & 54 & 5.5 & 2.78 & 3.34 \\\hline
RCA-200-20 & 469 & 224 & 5.5 & 1.41 & 2.10 \\\hline
RCA-300-30 & 410 & 204 & 5.5 & 2.00 & 3.30 \\\hline
QFT-100-10 & 2068 & 1784 & 18 & 8.70 & 6.53 \\\hline
QFT-200-20 & 8351 & 7566 & 18 & 9.10 & 6.98 \\\hline
QFT-300-30 & 18835 & 17348 & 18 & 9.24 & 7.13 \\\hline
BV-100-10 & 9 & 0 & 8 & 6.22 & 4.33 \\\hline
BV-200-20 & 19 & 0 & 8 & 6.63 & 4.63 \\\hline
BV-300-30 & 29 & 0 & 8 & 6.69 & 4.69 \\\hline
QAOA-100-10 & 1448 & 266 & 6 & 2.17 & 1.83 \\\hline
QAOA-200-20 & 6787 & 728 & 8 & 2.07 & 1.79 \\\hline
QAOA-300-30 & 16053 & 1138 & 6 & 2.05 & 1.69 \\\hline
UCCSD-8-4 & 464 & 0 & 4 & 1.94 & 1.74 \\\hline
UCCSD-12-6 & 8973 & 0 & 4 & 1.69 & 1.55 \\\hline
UCCSD-16-8 & 33303 & 0 & 5 & 1.60 & 1.50 \\\hline
\end{tabular}
    }
    \caption{The results of {\frameworkName} and its relative performance to the baseline. The name column are acronyms of test programs in Table~\ref{tab:benchmark}.}
    \label{tab:comm-cost}
\end{table}

\subsection{Compared to GP-based Compiler}

We further compare {\frameworkName} to the graph-partition-based (GP-based) compiler~\cite{time-slice}. 
A GP-based compiler converts remote interactions to local interactions by swapping qubits with a strategy derived from graph partition algorithms.
To reduce the communication cost and program latency of the GP-based compiler, we utilize TP-Comm for swapping qubits since TP-Comm requires only two communications for one remote SWAP gate, one communication less than using Cat-Comm. We denote this version of the GP-based compiler by \textit{GP-TP}. Once again, for GP-TP, we adopt the as-soon-as-possible schedule strategy in~\cite{Ferrari2021CompilerDF}.

\begin{figure}[tb]
\centering
    \includegraphics[width=0.40\textwidth]{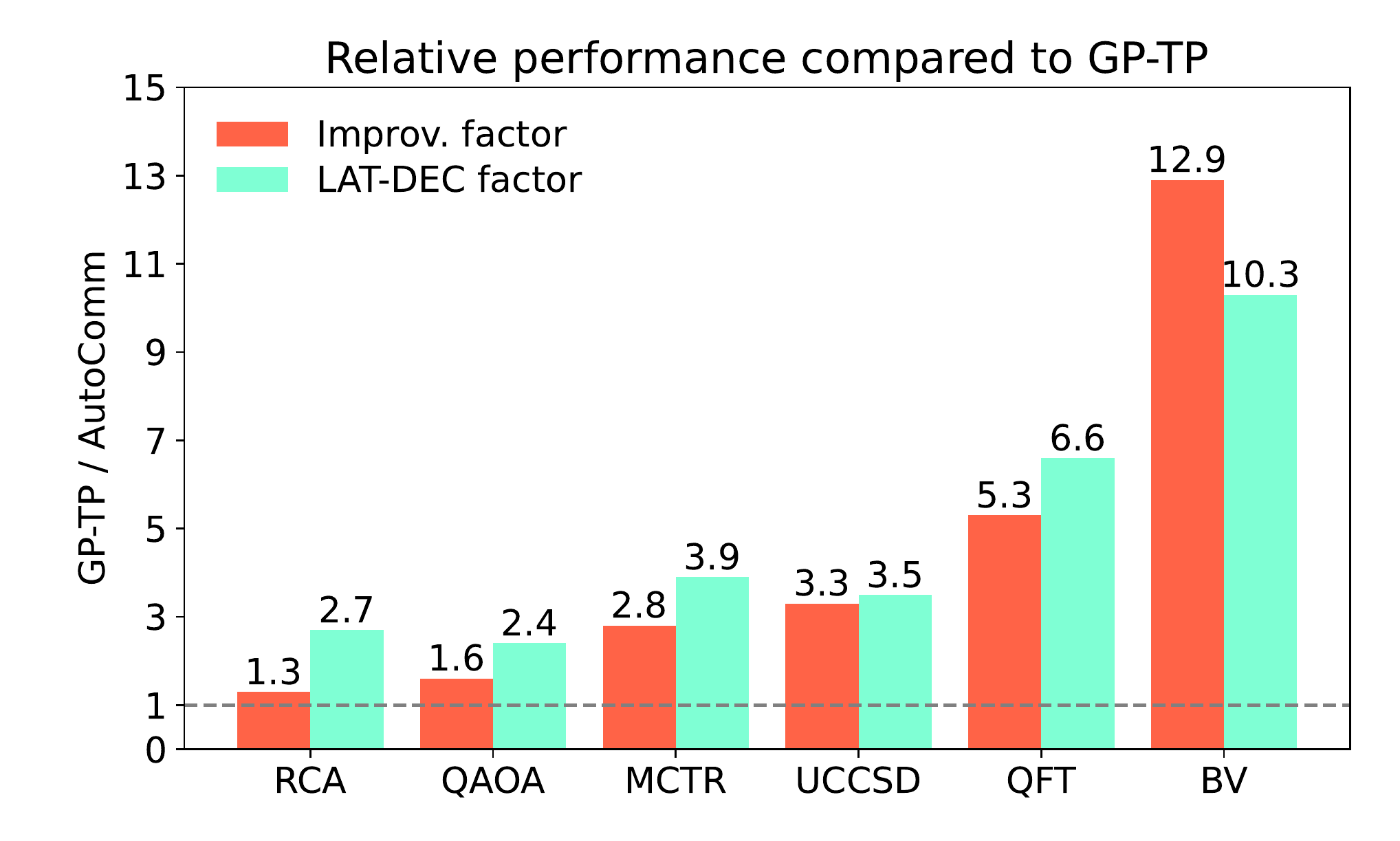}
    \caption{Compared to GP-TP. Results are averaged over different configurations of \#\,qubit and \#\,node in Table~\ref{tab:benchmark}.}
    \label{fig:gptp}
\end{figure}

As shown in Figure~\ref{fig:gptp}, {\frameworkName} achieves significant reduction in both communication cost and program latency, compared to GP-TP. Specifically, {\frameworkName} reduces the communication cost by a factor of 3.3x on average, up to 12.9x. It also reduces the program execution time by a factor of 4.3x on average, up to 10.3x. On the side of information theory, {\frameworkName} improves the performance by enabling  a higher throughput of information. Each remote communication in GP-TP carries less than two remote CX gates which is much smaller than {\frameworkName}. On the algorithmic side, {\frameworkName} reduces unnecessary qubit movement by taking advantage of burst communication. For example, for a potential burst communication between $q_1$ and node B, if there are some commutable remote CX gates between $q_1$ and node C lying in between and interrupting the communication block between $q_1$ and node B, the GP-TP method needs to move $q_1$ to node B first, then to node C and back to node B again. However, with burst communication, we only need to first move $q_1$ to node B, and then to node C.

\subsection{Optimization Analysis}\label{sect:optianaly}

\begin{figure*}[tb]
\newcommand{\fhe}{0.142}
    \includegraphics[height=\fhe\textwidth]{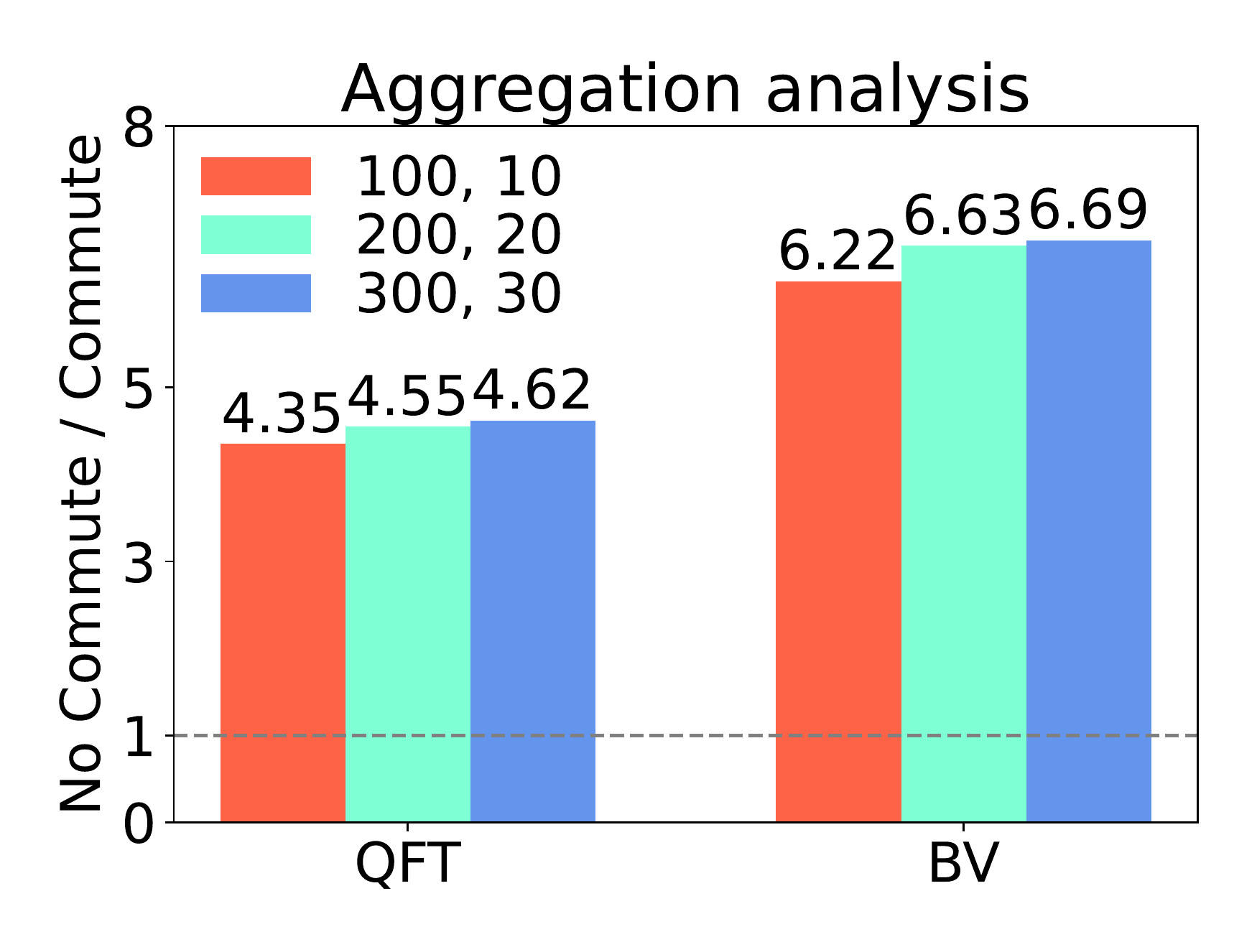} \hfill \includegraphics[height=\fhe\textwidth]{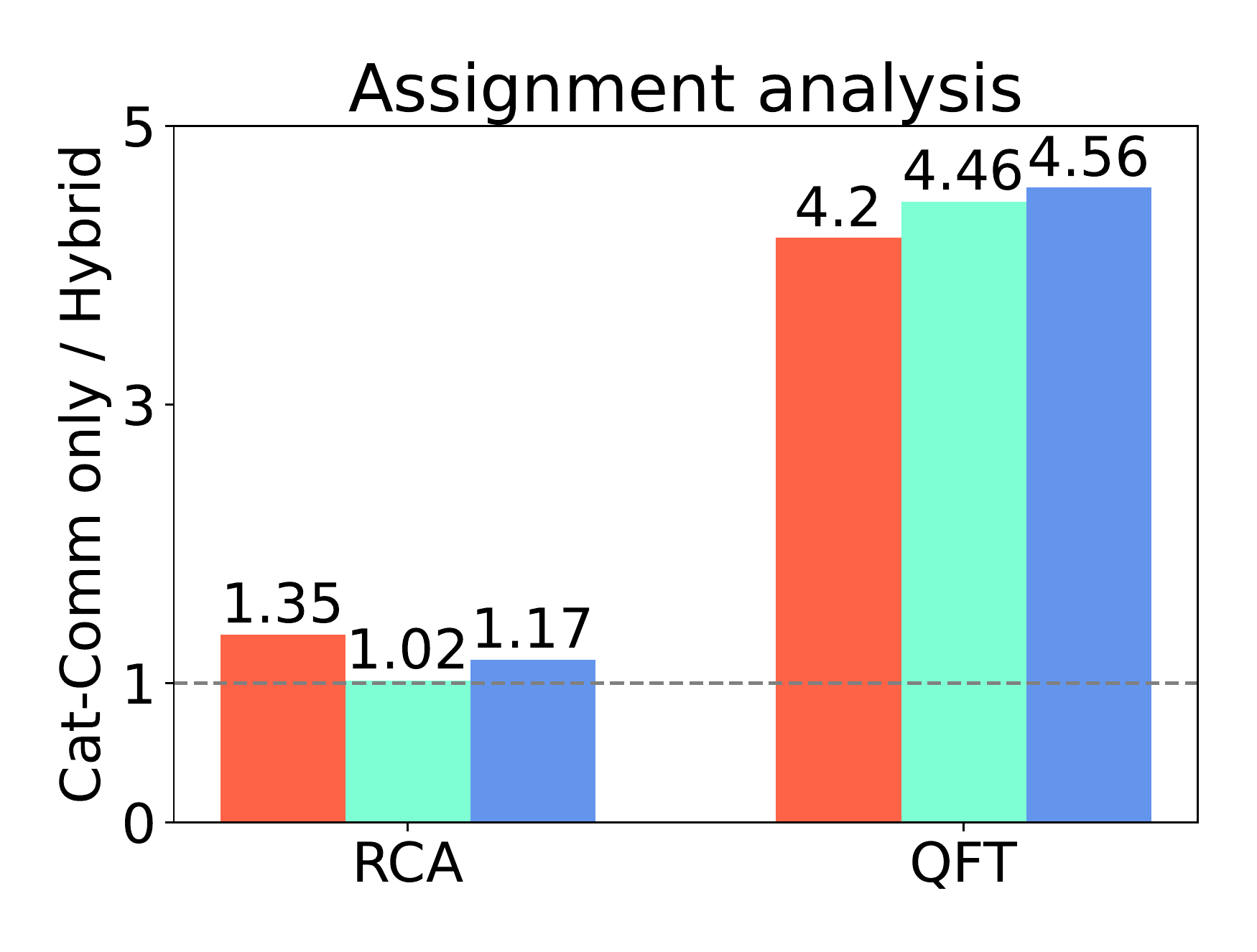} \hfill \includegraphics[height=\fhe\textwidth]{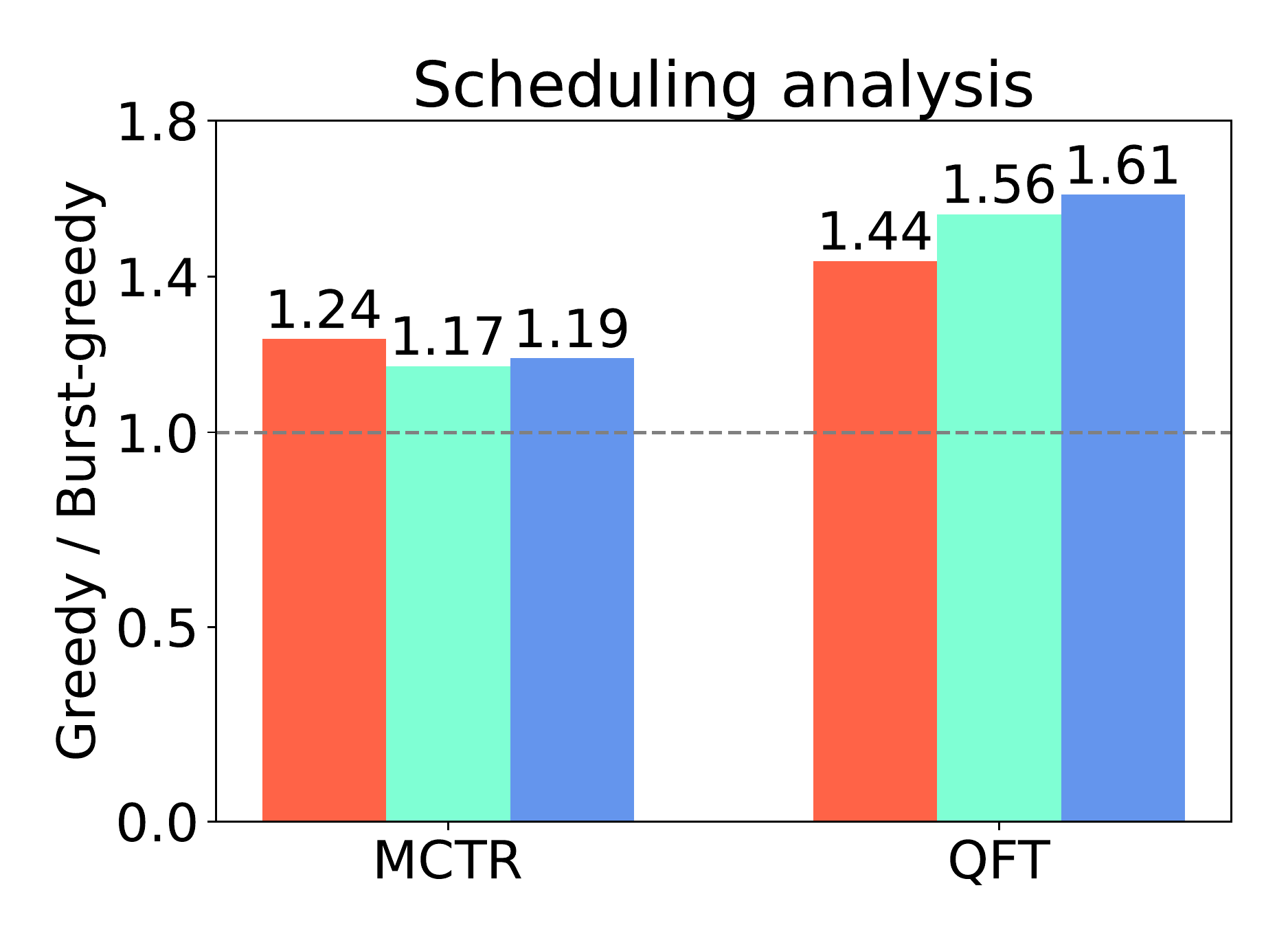} \hfill
    \includegraphics[height=\fhe\textwidth]{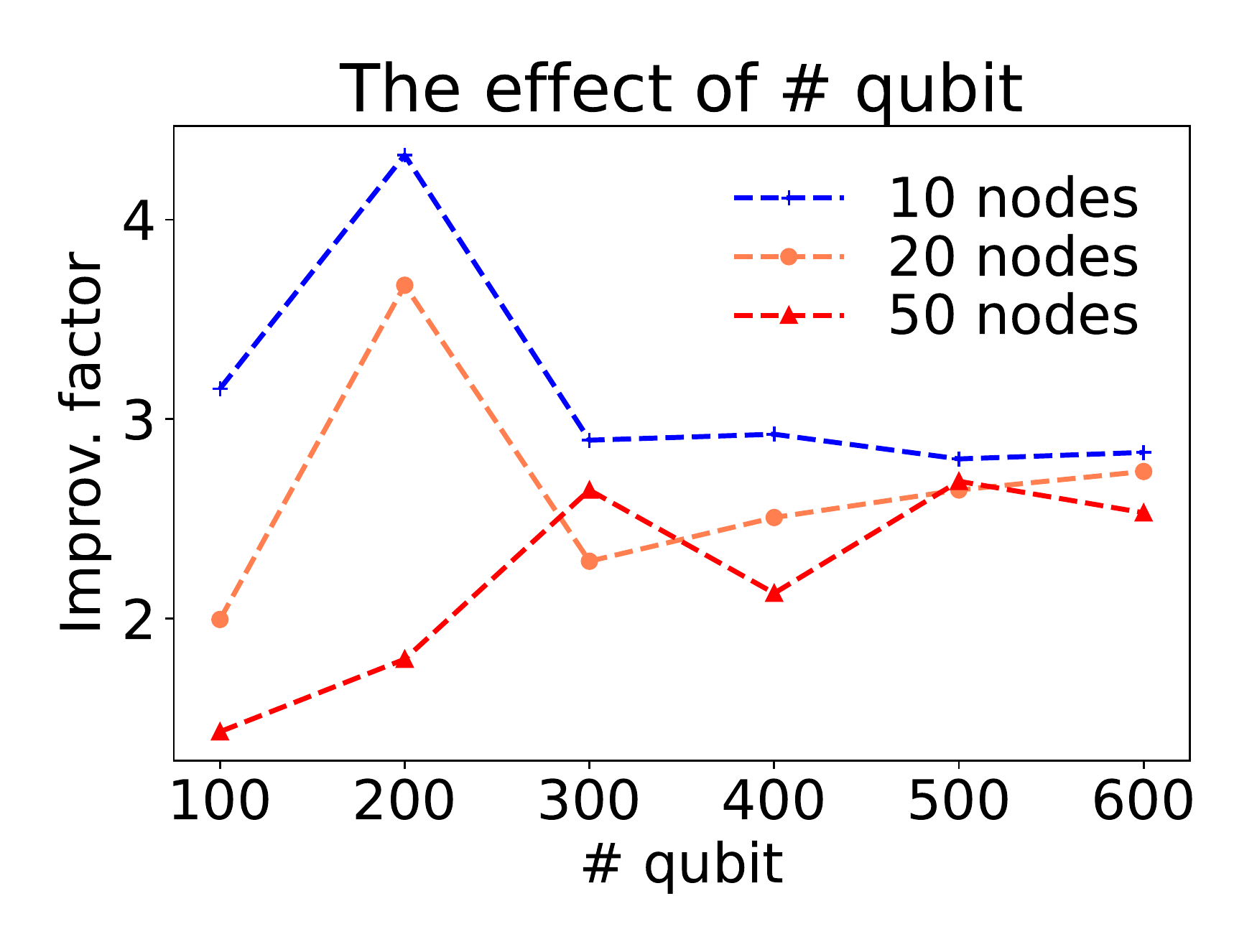} \hfill
    \includegraphics[height=\fhe\textwidth]{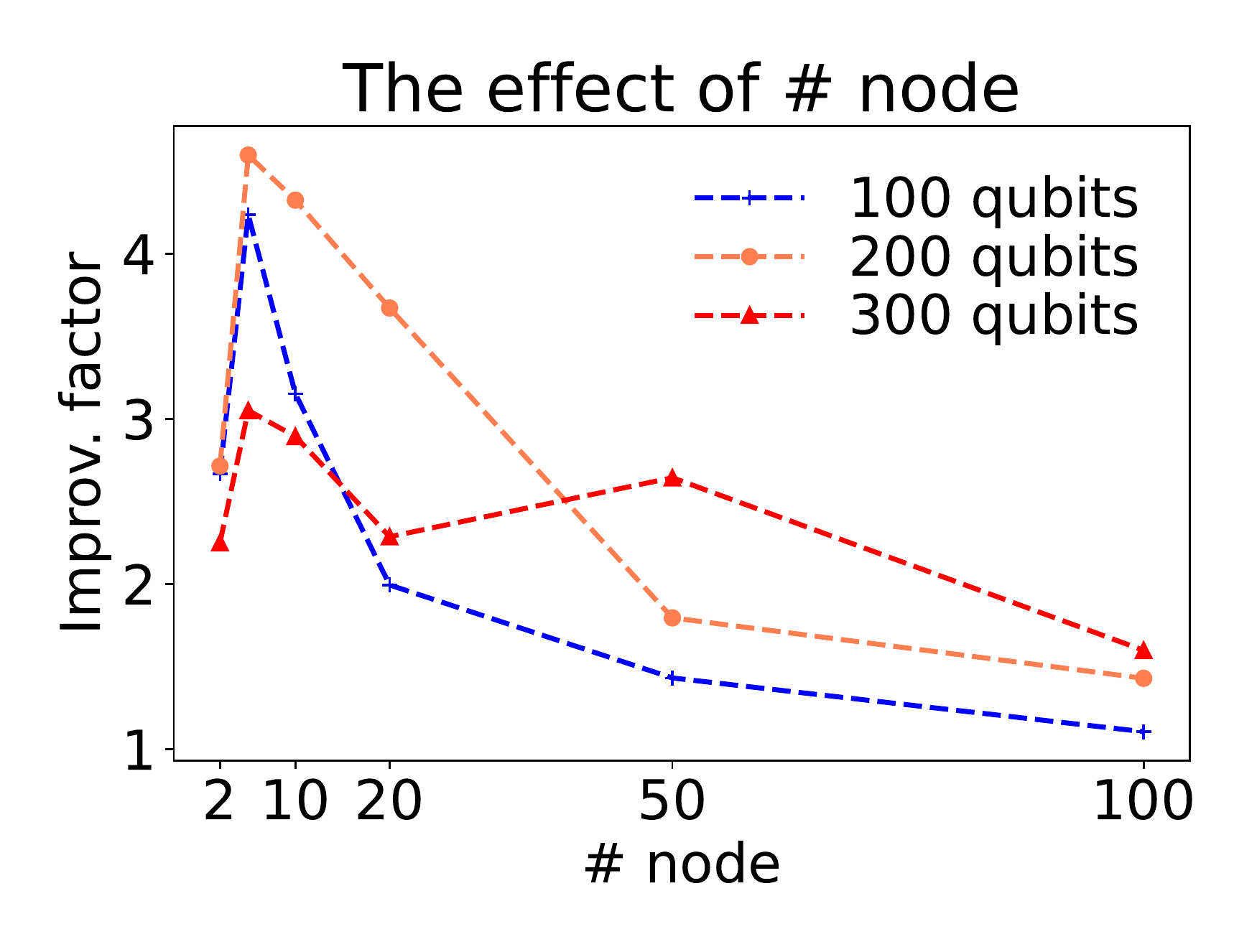} \\
    \hphantom{} \hspace{48pt}(a) \hspace{85pt} (b) \hspace{85pt} (c) \hspace{85pt} (d) \hspace{85pt} (e)
    \caption{(a)-(c) The effects of the proposed optimizations. Bars with different colors denote different configurations of (\#qubit, \#node). %
    For (a)(b), the y-axis is the ratio of \# remote communications. For (c), the y-axis is the ratio of program latency. (d)(e) The effects of \#\,qubit and \#\,node on the improv. factor of \frameworkName. The test program in (d)(e) is MCTR.}
    \label{fig:opti}
\end{figure*}

In this section, we further explore and analyze the effect of each optimization in \frameworkName. For each analysis, we change only one component of {\frameworkName} at a time, with other components fixed, to isolate the effect of each component/optimization.

\paragraph{The effect of communication aggregation:} Table~\ref{tab:comm-cost} demonstrates the benefit of communication aggregation compared to the baseline. Here we further demonstrate the necessity of considering gate commutation in the aggregation pass. Figure~\ref{fig:opti}(a) shows the communication cost comparison between the aggregation without gate commutation and the aggregation used in {\frameworkName}. For the programs in Figure~\ref{fig:opti}(a), {\frameworkName} reduces the communication cost by a factor of 5.5x on average, up to 6.7x, compared to the aggregation without gate commutation.
Gate commutation is indispensable for discovering burst communications, not only because multi-qubit gates are often scattered in quantum circuits, but also due to the uncertainty of qubit mapping to quantum nodes (the uncertainty of whether a CX is remote or local).

\paragraph{The effect of hybrid communication assignment:} We further demonstrate the importance of considering both Cat-Comm and TP-Comm for burst communication. Figure~\ref{fig:opti}(b) shows the communication cost comparison between the communication assignment with Cat-Comm only and the hybrid assignment scheme in {\frameworkName}. The Cat-Comm only method is extended from the specialized compiler~\cite{Diadamo2021DistributedQC} for distributed VQE.
For the programs in Figure~\ref{fig:opti}(b), {\frameworkName} reduces the communication cost by a factor of 2.8x on average, up to 4.6x, compared to the Cat-Comm only method. The key enabler for the hybrid scheme in {\frameworkName} is that Cat-Comm only applies to few communication patterns and for the cases that Cat-Comm cannot apply, TP-Comm would be more efficient. %

\paragraph{The effect of communication scheduling:} We then study the effect of the communication scheduling optimization in {\frameworkName}. %
Figure~\ref{fig:opti}(c) shows the latency comparison between {\frameworkName}'s scheduling, denoted by burst-greedy, and the greedy (as-soon-as-possible) scheduling for communication blocks.
For the programs in Figure~\ref{fig:opti}(c), the burst-greedy method reduces the program latency by a factor of 1.4x on average, up to 1.6x, compared to the general greedy schedule. 
The effectiveness of {\frameworkName} for scheduling burst communication stems from its smart utilization of communication qubits, especially for TP-Comm blocks, as discussed in Section~\ref{sect:schedule}.

\subsection{Sensitivity Analysis}

The performance of {\frameworkName} may be affected by some external factors, e.g., the number of input qubits and the number of computing nodes.
In this section, we study how the performance of {\frameworkName} varies with those factors. We focus on `improv. factor' here, and the variation of `LAT-DEC factor' would follow a similar trend.

\paragraph{The effect of \#\,qubit:} Figure~\ref{fig:opti}(d) shows how the improv. factor of {\frameworkName} changes with the number of qubits. As shown in the figure, the improv. factor converges when \# qubit/\# node is large. 
This may be due to the fact that the number of burst communication blocks also increases when the total number of remote multi-qubit gates grows with the number of qubits. Such behavior is preferable because it illustrates that {\frameworkName} can provide a consistent reduction for the communication cost as the number of qubits grows.

\paragraph{The effect of \#\,node:} Figure~\ref{fig:opti}(e) shows how the improv. factor of {\frameworkName} changes with the number of nodes. In this figure, the performance of {\frameworkName} deteriorates when \# qubit/\# node is small. This is because it is harder to find large communication blocks when the number of qubits in each node is limited to be small.
Therefore we should avoid using too many nodes for distributing quantum programs because in such a case the remote multi-qubit gates would proliferate and there is little chance to execute those remote interactions collectively, given the fact that the number of communication qubits in each node is only two.

%% file: 07_discussion.tex
\section{Discussion and Future Work}

To the best of our knowledge, this paper is the first attempt that formalizes burst communication in distributed quantum programs. 
We discover a large number of burst communications hidden in various distributed quantum programs and propose the first modular framework to uncover these burst communications and use them to optimize the communication overhead. 
We argue that the formalization of burst communication and the modular solution proposed in this paper unveil new opportunities for communication optimization in DQC and would potentially inspire a series of works for overcoming DQC's communication problem.

Although we show that the proposed framework significantly surpasses existing works in optimizing the communication overhead of distributed quantum programs, there is still much space left for potential improvements. 

\paragraph{Extending to general collective communication} This paper only considers the near-term DQC where communication qubits are supposed to be limited. In such a  case, we are restricted to studying the qubit-to-node burst communication, which is a special case of the general collective communication, involving a group of nodes. 
Assuming the availability of more communication qubits in the future, we could consider node-to-node collective communication which offers a potential optimization opportunity as we can now aggregate small qubit-to-node burst communication blocks into a larger one. Besides, for the fusion operation in the communication schedule optimization, we can also extend it to node-to-node communication blocks.

\paragraph{Co-designing with front compiling stages} The proposed framework is designed to be easily pluggable into existing compiling flows. But we could also couple it with front compiling stages to achieve further optimization. For example, 
existing compilers include a pass to add SWAP gates to change the qubit layout to optimize circuit metrics. We could co-design with this pass to maximize the number and  size of burst communications. Besides, in the case where burst communication is deeply hidden, we could also consider using unitary synthesis to create burst communication in the gate decomposition pass. %
Finally, we could co-design with the qubit mapping pass to achieve a balance of communication overhead and device utilization rate, as shown in Figure~\ref{fig:opti}(d)(e).

\paragraph{Combining with quantum error correction} Since DQC involves quantum communication which is far more noisy than local quantum gates, reinforcing the whole distributed quantum system with quantum error correction (QEC) becomes vital for future DQC. 
One promising way to implement QEC in DQC is to encode one logical qubit in each node, and use quantum communication to implement logical operations between logical qubits. In this case, the CX gate between logical qubits would involve a large number of physical qubits simultaneously and provide great opportunities for burst communication optimization. Besides, communications coming from magic state distillation are also worth considering.

%% file: 03_related_work.tex
\section{Related Work} 

Most existing quantum compilers~\cite{Li2019TacklingTQ, Qiskit, Amy2019staqAFQ, Khammassi2022OpenQLA, Sivarajah2020tketAR} focus on the compilation of programs within a single quantum computer. Extending these works to DQC cannot achieve high information throughput per quantum communication, as in the compiler proposed by Ferrari et al.~\cite{Ferrari2021CompilerDF}. Baker et al.~\cite{time-slice} propose using the more informative remote SWAP gates to replace all remote CX gates in distributed quantum programs and obtain a higher throughput. Diadamo et al. \cite{Diadamo2021DistributedQC} further increase the communication throughput by considering multiple-qubit control-unitary blocks. However, their work requires specialized circuit representation and cannot optimize general quantum programs. 
Moreover, all these works do not consider the burst communication and related optimizations proposed in this paper.

Another line of work executes distributed quantum programs in a hybrid way. Tang et al.\cite{cutQC} propose a way to execute quantum programs in distributed computing nodes but without inter-node communication. To overcome the expressibility loss due to no inter-node communication, 
their work relies heavily on classical post-processing techniques and cannot be extended to large-scale quantum programs.

Other quantum communication-related works focus on building robust quantum communication networks~\cite{QNrouting1,QNrouting2,QNrouting3,QNrouting4} or reducing the resource consumption of existing quantum communication techniques~\cite{repeater12,repeater13,repeater14,repeater15,repeater16}. These works are orthogonal to this paper.

%% file: 08_conclusion.tex
\section{Conclusion}

As in classical distributed computing, the inter-node communication overhead bottlenecks distributed quantum computing. Existing compilers~\cite{Ferrari2021CompilerDF, Diadamo2021DistributedQC, time-slice} for distributed programs either treat the inter-node communication like the in-node communication or only provide optimization for gates in the control-unitary form. These works fail to utilize the hidden communication patterns in distributed quantum programs. To overcome the shortcomings of existing compilers, this paper explores various distributed quantum programs and identifies burst communication for the first time. Burst communication is a qubit-node communication pattern that widely exists in many distributed programs. 
Based on burst communication, we propose the framework, {\frameworkName}, which is proved to be efficient in cutting down inter-node communication overhead, by comprehensive evaluations on diverse distributed benchmarks.
The proposed framework can be easily integrated into existing compiling flows of quantum programs and would benefit near-term distributed quantum computing.